\documentclass[bibyear]{aa}  

\usepackage{xcolor}
\usepackage{bbm} 
\usepackage{mathtools}
\usepackage{natbib}
\usepackage{hyperref}
\usepackage{silence}
\usepackage{microtype}
\usepackage{ulem}
\usepackage{graphicx}
\usepackage{csquotes}
\usepackage{needspace}
\usepackage{txfonts}
\usepackage{makecell}
\usepackage[caption=false]{subfig}
\usepackage{booktabs}  
\usepackage{tabularx} 
\makeatletter 
\newcommand{\@giventhatstar}[2]{\ensuremath{\left({#1}\,\middle|\,{#2}\right)}} 
\newcommand{\@giventhatnostar}[3][]{#1(#2\,#1|\,#3#1)} 
\newcommand{\giventhat}{\@ifstar\@giventhatstar\@giventhatnostar} 
\makeatother

\makeatletter 
\newcommand{\@giventhatstardouble}[2]{\ensuremath{\left({#1}\,\middle|\middle|\,{#2}\right)}} 
\newcommand{\@giventhatnostardouble}[3][]{#1(#2\,#1||\,#3#1)} 
\newcommand{\giventhatdouble}{\@ifstar\@giventhatstardouble\@giventhatnostardouble} 
\makeatother

\DeclarePairedDelimiter\abs{\lvert}{\rvert}%
\DeclarePairedDelimiter\norm{\lVert}{\rVert}%

\makeatletter
\let\oldabs\abs
\def\abs{\@ifstar{\oldabs}{\oldabs*}}
\let\oldnorm\norm
\def\norm{\@ifstar{\oldnorm}{\oldnorm*}}
\makeatother

\begin{document} 

\defcitealias{Suzuki_2012}{Union2.1}
\defcitealias{Pantheon_2022}{Pantheon+}
\defcitealias{DESY5}{DESY5}

   \title{Nonparametric Variational Inference Reconstruction of the Cosmic Expansion History from SNe Ia - the \texttt{charm2} code}


   \author{Iason Saganas
          \inst{1,2}, 
          Matteo Guardiani \inst{1, 2},
          Natalia Porqueres
          \inst{3}
          \and
          Torsten Enßlin \inst{1, 2, 4, 5}
          }

    \institute{Ludwig-Maximilians-Universit\"at M\"unchen, Geschwister-Scholl-Platz 1,
80539 Munich, Germany\\
              \email{isaganas@mpa-garching.mpg.de}
         \and
             Max Planck Institute for Astrophysics, Karl-Schwarzschildstr. 1, 85748 Garching, Germany.
         \and
            Université Paris-Saclay, Université Paris Cité, CEA, CNRS, AIM, 91191, Gif-sur-Yvette, France
        \and 
        Deutsches Zentrum für Astrophysik, Postplatz 1, 02826 Görlitz, Germany
        \and 
        Excellence Cluster ORIGINS, Boltzmannstr. 2, 85748 Garching, Germany
            \\ 
             }

   \date{Received here; accepted then}

 
  \abstract
   {
   Cosmological analyses using the latest set of type Ia Supernova data weakly favor an evolving dark energy (EDE) model without strongly disfavoring the standard $\Lambda$CDM paradigm. Nonparametric reconstructions of the expansion history may reveal signal features potentially missed by a parametric $\Lambda$CDM model without laying out a specific functional form for the evolution of dark energy. Information field theory (IFT) is a Bayesian framework for optimal, nonparametric reconstruction algorithms.}
   {In this work, we present \texttt{charm2}, the successor to \texttt{charm1}, a previous IFT-based code to reconstruct the cosmic energy density's redshift evolution from Supernovae Ia. 
   We apply our reconstruction algorithm to the Union2.1, Pantheon+, DESY5 and DESY5-Dovekie data sets to investigate the agreement between the nonparametric reconstruction and the signal suggested by a parametric, flat $\Lambda$CDM model.}
   {To enable an accurate Gaussian approximation, we employ geometric variational inference, which finds a coordinate transformation through which a curved posterior gets "flattened". The redshift evolution of the energy density can then be traced on a double-logarithmic scale, which, after de-trending, is well described by a stationary Gaussian process. 
}
   {The nonparametric $\texttt{charm2}$ reconstructions using the Union2.1 and Pantheon+ data sets are consistent with flat $\Lambda$CDM signal fields.
   The DESY5 and DESY5-Dovekie reconstructions deviate from flat $\Lambda$CDM comparison fields and are compatible with an evolving dark energy signal. However, using the evidence lower bound (ELBO) measure for model selection, we find no conclusive evidence supporting a preference for non-flat-$\Lambda\mathrm{CDM}$ features in any of the data sets. 
   We note that at current DESY5 noise levels, the ELBO tends to favor flat $\Lambda\mathrm{CDM}$ over our nonparametric model, although the latter better recovers the ground truth in synthetic EDE data; a trend reversing only at $\sim 7\times $ lower noise covariance.
   
}
   {}


    \titlerunning{Cosmic expansion history from SNe Ia data – the \texttt{charm2} code}
    \authorrunning{Saganas et al.}
   \maketitle
%

\section{Introduction}

The detection of the universe's accelerated expansion by \cite{Perlmutter_1999} implied that most of the universe's energy content is due to an unknown energy density, called dark energy, which appears to remain constant even as the universe expands. Together with observations providing strong evidence for the existence of a cold dark matter component (such as the weak lensing map of the bullet cluster, rotation curves of galaxies, etc, see \cite{Freese_2009} for a review), the so-called $\Lambda$CDM model has been the favored candidate for a standard model of cosmology. Cosmological analyses based on this model revealed that only about $5\%$ of the universe's energy content is baryonic, with approximately $95\%$ bound up in the dark matter and dark energy components \citep{Freese_2009}.

In recent years, models that allow the dark energy density, $\rho_{\Lambda}$, to evolve over cosmic time have regained traction, with recent analyses of the Dark Energy Spectroscopic Instrument favoring a dynamical dark energy scenario \citep{Adame_2025}. Often, a parametric form is assumed for the evolution of the dark energy component. Cosmological parameters, such as the relative energy density fractions in the universe, can then be extracted by minimizing the $\chi^2$ between the predicted and observed data with respect to the parameters of interest. If the assumed model, whether the $\Lambda$CDM model or a parametric evolution of dark energy, is incorrect, the analysis results will be biased.

To circumvent such model bias, different methods have been developed to extract cosmological information from observed data without assuming a physical model ("agnostic inference") or relying on a parametric form ("nonparametric inference"). 
Usual nonparametric reconstruction methods are Gaussian Processes \citep{GP1, GP6, GP2, GP3, Spl2_GP4, GP5}, Principal Component Analyses (PCA) \citep{PCA2, PCA1}, Splines and other polynomial-based approaches (for example Padé-Polynomials) \citep{Spl1, Pade1, Spl2_GP4}, as well as smoothing approaches \citep{Smoothing1, Smoothing2}. In recent years, Neural-Network-based reconstructions have emerged as well \citep{NN2, NN1}.  

\cite{Porqueres_2017} proposed a nonparametric and agnostic approach to reconstruct the evolution of the cosmic expansion in the late universe through Supernovae Ia (SNIa) inside the framework of information field theory (IFT) \citep{Ensslin_2009_ift}. Their approach relied on a Wiener filter and a linearized response operator through which the posterior distribution was obtained by recursively perturbing the solution and reapplying the Wiener filter until a stationary solution is reached. This describes an iterative maximum-a-posteriori scheme in which the posterior mode is approximated by a Gaussian.

Here, we improve this approach by using the geometric variational inference algorithm \citep[\texttt{geoVI,}][]{Frank_2021} to account for the nonlinear terms of the data model and to provide more accurate posterior uncertainties.

Rather than defining constant piecewise functions  \citep{PCA_for_EoS} or polynomials \citep{PCA1}, our method does not rely on any functional basis and further accounts for the full Fisher information metric, as opposed to PCA approaches, which often truncate the eigenvalues of the Fisher information metric. As in neural network frameworks (e.g., \cite{NN1}), our approach introduces hyperparameters but does not require training datasets.

In this paper, we present $\texttt{charm2}$ (cosmic history agnostic reconstruction method No. 2), a code for nonparametrically reconstructing the redshift evolution of the universe's energy density, with improved treatment of nonlinearities and uncertainties when compared to its predecessor in \cite{Porqueres_2017}. \texttt{charm2} is built using the Python package \texttt{NIFTy}\footnote{\texttt{NIFTy} (numerical information field theory) is available at \href{https://ift.pages.mpcdf.de/nifty/user/index.html}{https://ift.pages.mpcdf.de/nifty/user/index.html}}. 

The paper is structured as follows: in Sect. \ref{sec: Inference Approach}, we introduce the concepts necessary to set up our agnostic inference approach of the cosmic expansion history, defining the inferred signal and the data model. We describe the data sets analyzed in this work in Sect. \ref{sec: Data}. Finally, we present our analysis of the Union2.1, Pantheon+, DESY5 and DESY5-Dovekie data sets in Sect. \ref{sec: Results} before summarizing our findings in Sect. \ref{sec: Conclusions}.

\begin{figure*}
   \centering
   \includegraphics[width=0.9\linewidth]{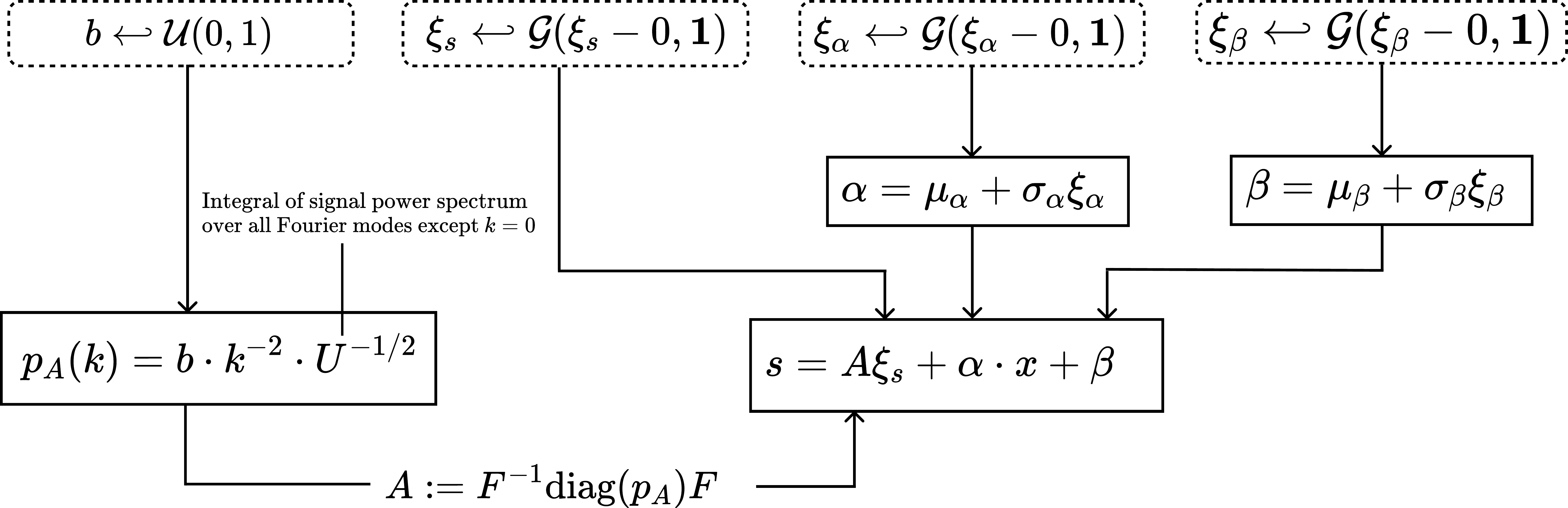}
   \caption{The generative model used for the inference of the cosmic expansion history as encoded in $\mathrm{ln}(\rho/\rho_0)$ in \texttt{charm2}. The right side of the graph describes the generation of the linear part of the signal $s$ in \texttt{charm2}'s model. The linearly evolving part of the signal, $\alpha x+\beta$, captures the dominant part of the density evolution and corresponds to the dominant cosmic constituent. The left-hand side of the graph describes the generation of the nonparametric $A\xi_s$ part of the signal, which accounts for deviations from a single-component equation of state.
   The random variates $\alpha$, $\beta$, $n$ and $b$ are described by the distributions shown in Table \ref{table: Hyperparameters chosen for the generative prior model}.}
              \label{fig: Computational graph of Correlated Field Model}%
\end{figure*}


\section{Inference approach}\label{sec: Inference Approach}

\subsection{Basic theory and signal definition}\label{subsec: Basic theory and signal definition}

Conceptually, the standard $\Lambda$CDM model of cosmology rests upon the cosmological principle, the notion that matter is distributed homogeneously and isotropically on large scales. On these scales, matter is described as a perfect fluid following a simple equation of state (EOS),
\begin{equation}\label{eq: Equation of state imposed on perfect fluid}
    p_i=w_i\rho_i,
\end{equation}
where $\rho_i$, $p_i$ and $w_i$ are the energy density, pressure, and EOS parameter, respectively.
In $\Lambda$CDM, the EOS parameter $w$ is assumed to be a constant for each fluid $i$ contributing to the universe's total energy density $\rho$:
\begin{equation}\label{eq: Total energy density as sum over individual densities}
    \rho = \sum_i \rho_i = \sum_i \rho_{0i}a^{-3(1+w_i)}.
\end{equation}
Here, $\rho_{0i}$ is the energy density of the $i$-th contribution today and $a$ is the scale factor related to the redshift $z$ through $a=(1+z)^{-1}$. Contributions may include ordinary and dark matter, radiation, dark energy, and the universe's intrinsic curvature, denoted by $\rho_m$, $\rho_r$, $\rho_{\Lambda}$, and $\rho_k$, respectively. 

One may then solve the Einstein field equations to determine the universe's evolution through its expansion rate, quantified by the Hubble parameter
\begin{equation}
    H(a)=\frac{\dot{a}}{a}.
\end{equation}
The expansion rate is related to the universe's total energy density via the first Friedmann equation:
\begin{equation}\label{eq: Friedmann equation in the LCDM model}
    H(a)^2 = \frac{8\pi G}{3}\rho =  \frac{8\pi G}{3} \bigg(\rho_{0m}a^{-3}+\rho_{0r}a^{-4}+\rho_{0k}a^{-2}+\rho_{0\Lambda}\bigg).
\end{equation}
Here, the dependence on the scale factor $a$ is characteristic for each type of fluid, arising from different values of the EOS parameter $w$ (e.g., $w=0$ for matter, $w=1/3$ for radiation and $w=-1$ for a constant dark energy). At a given redshift, there is usually one dominant contribution to the universe's total energy budget, defining different cosmological eras.

In this work, we do not assume specific contributions to the total energy density $\rho$ as the number of contributing components and their evolution through cosmic time is not known a-priori. Instead, we generally define the signal $s$ as the energy density on a double-logarithmic scale
\begin{equation}\label{eq: Signal definition v2}
    s(x):=\mathrm{ln}\bigg(\frac{\rho (x)}{\rho_0}\bigg),
\end{equation}
where $x=-\mathrm{ln}(a)=\mathrm{ln}(1+z)$ and $\rho_0$ is an arbitrary reference energy density. 

For $\Lambda$CDM, the signal field $s_{\mathrm{\Lambda CDM}}(x)$ takes on the form of a piecewise linear function in these coordinates (e.g., in the radiation-dominated era, $s_{\mathrm{\Lambda CDM}}(x) \propto -4\: \mathrm{ln}(a)=4x$). Each era corresponds to a linear function with a slope proportional to its EOS parameter, smoothly connecting to the linear function representing the next era.

Motivated by the piecewise linearity of the $\Lambda\mathrm{CDM}$ model in these coordinates, we model $s(x)$ through a nonparametric, stationary process $s_\mathrm{dev}(x)$ around a mean linear trend $s_\mathrm{lin}(x)$ (see Sect. \ref{subsection: Generative model prior}), by imposing homogeneous statistics on the deviations $s_\mathrm{dev}(x)$, i.e. imposing that the covariance $S^\mathrm{dev}$ at two points $x$ and $y$ only depends on their difference:
\begin{equation}\label{eq: Stationary covariance matrix}
    S^\mathrm{dev}_{xy} = C_s(x-y),
\end{equation}
with $C_s$ being the $2$-point-correlation function. This is justified because the nonparametric component only needs to capture oscillations around the mean linear trend. In practice, thanks to the log-log definition of the signal (Eq. \eqref{eq: Signal definition v2}), the amplitude of these oscillations varies by roughly one order of magnitude at most. Therefore, a stationary process is well suited, since the variance $S^\mathrm{dev}_{xx}=C_s(0)$ is constant for all $x$.

\subsection{Instrument response}\label{subsec: Instrument response}

To constrain the expansion history of the universe, one measures the redshifts of objects at cosmological distances. The way redshift varies with distance encodes information about how the universe has expanded over time. 

The luminosity distance provides the distance to an object based on the observed flux it emits. In a flat universe and assuming the FLRW metric, the luminosity distance is given by
\begin{equation}\label{eq: Luminosity distance in a flat FLRW universe}
    d_L (z)=(1+z) \int_0^{z} \frac{c}{H(z')}  \:\mathrm{d}z'.
\end{equation}
Through a multi-step calibration process known as the cosmic distance ladder, the luminosity distance is related to the observed distance modulus of SNIa via
\begin{equation}\label{eq: Distance modulus for SN}
    \mu = 5\mathrm{log}_{10}\big(d_L^{\mathrm{(pc)}}\big) -5,
\end{equation}
where $d_L^{\mathrm{(pc)}}$ denotes the luminosity distance in units of $\mathrm{pc}$. These distance moduli constitute our discrete set of data $d$, which depend on the physical quantity $s=\mathrm{log}(\rho/\rho_0)$ we want to infer: the energy density determines the evolution of the Hubble parameter through the first Friedmann equation. The Hubble parameter in turn influences the luminosity distance via Eq. \eqref{eq: Luminosity distance in a flat FLRW universe}. 

Through our signal definition Eq. \eqref{eq: Signal definition v2}, the Hubble parameter can be expressed as  
\begin{equation}\label{eq: First Friedmann Equation}
    H^2(x)=\frac{8\pi G}{3}\rho(x) =\frac{8\pi G}{3} \rho_0 e^{s(x)}.
\end{equation}
We set the reference energy density $\rho_0$ to 
\begin{equation}\label{eq: energy density normalization}
\rho_0 = 10^6 \frac{\mathrm{kg}}{\mathrm{m\cdot Mpc}^2},
\end{equation}
such that the Hubble-Parameter is in the usual units of $\mathrm{km/s/Mpc}$: $[H(a)]=\sqrt{[G]\cdot [\rho]}=\sqrt{[G]\cdot \rho_0}=\mathrm{km/s/Mpc}$, where $G$ is Newton's gravitational constant in SI-units.

In IFT, the function that maps the signal to the (noiseless) data is defined as the response operator. This function encapsulates the complex physical processes leading to the data generation. Using the SNIa distance modulus, Eq. \eqref{eq: Distance modulus for SN}, together with Eq. \eqref{eq: First Friedmann Equation} allows us to write the response operator as
\begin{equation}\label{eq: Instrument Response Operator}
    R_x(s)=5\mathrm{log}_{10}\left(e^x \mathcal{K} \int_0^x e^{-\frac{1}{2}s({\tilde{x}})+\tilde{x}}\hspace{1mm}\mathrm{d}\tilde{x}\right) -5,
\end{equation}
where $\mathcal{K}$ is a constant defined as
\begin{equation}
    \mathcal{K}=1000\cdot c\hspace{1mm}\bigg(\frac{8\pi G}{3}\bigg)^{-\frac{1}{2}},
\end{equation}
with $c$ and $G$ implicitly given in SI-units. The constant $\mathcal{K}$ is chosen such that the requirement of the luminosity distance to be in $\mathrm{pc}$, as expressed in Eq. \eqref{eq: Distance modulus for SN}, is met: the line-of-sight integral is implicitly an integral over $c/H(x)$ which is in $1000\cdot \mathrm{pc}$ since $H(x)$ is in $\mathrm{km/s/Mpc}$. Expressing $H(x)$ in terms of the signal field $s(x)$ through Eq. \eqref{eq: First Friedmann Equation} introduces the additional $8\pi G/3$ factor. We will write $\hat{H}_0$ to indicate a Hubble constant implicitly given in $\mathrm{km/s/Mpc}$, instead of writing out the unit explicitly.

Assuming additive noise, $n$, the data can be written as
\begin{equation}\label{eq: Instrument Response Equation}
    d=R(s)+n =: s^{\prime}+n,
\end{equation}
where $s^{\prime}$ is referred to as the noiseless data.

\subsection{Inference method}\label{subsec: Inference method}

Our inference method recovers the posterior distribution of the signal given the data, $\mathcal{P} \giventhat{s}{d}$, which is given by Bayes' Theorem as

\begin{equation}
   \mathcal{P} \giventhat{s}{d}=\frac{ \mathcal{P}\giventhat{d}{s}\mathcal{P}(s)}{\mathcal{P}(d)},
\end{equation}
where $\mathcal{P}\giventhat{d}{s}$ is the likelihood, $\mathcal{P}(s)$ is the signal prior and the normalisation $\mathcal{P}(d)$ is the evidence. In IFT, the evidence is redefined as a partition function,
\begin{equation}
    \mathcal{Z}(d) := \mathcal{P}(d)
\end{equation}
and negative log-probabilities as information Hamiltonians, 
\begin{equation}
\mathcal{H} = - \mathrm{ln}(\mathcal{P}).
\end{equation}
Then, Bayes' Theorem is given by
\begin{equation}\label{eq: Bayes Theorem in Hamiltonian/Boltzmann form}
\mathcal{P}\giventhat{s}{d} = \frac{1}{\mathcal{Z}}e^{-\mathcal{H}(s,d)}=\frac{1}{\mathcal{Z}}e^{-\mathcal{H}\giventhat{d}{s}-\mathcal{H}(s)}.
\end{equation}
To define the likelihood $\mathcal{H}\giventhat{d}{s}$, we assume a Gaussian distribution based on the Maximum Entropy principle. Therefore, $\mathcal{H}\giventhat{d}{s}$ is given by: 
\begin{equation}
    \mathcal{H} \giventhat{d}{s} \hspace{1mm}\widehat{=} \hspace{1mm}\frac{1}{2}(d-R(s))^{\dagger}N^{-1}(d-R(s)),
\end{equation}
where $\widehat{=}$ denotes equality up to constants. 

We assume that the matter distribution in the universe is homogeneous and isotropic on large scales and that the total energy density $\rho$ evolves smoothly with the scale factor, while it potentially may vary over multiple orders of magnitude. As discussed in Sect. \ref{subsec: Basic theory and signal definition}, the signal definition $s(x)=\mathrm{ln}(\rho(x)/\rho_0)$ with $x=-\mathrm{ln}(a)$ ensures that the amplitude of these variations around the mean linear trend is comparable among the different eras of cosmic history. A homogeneous process with respect to $x$ is then well-suited for these variations, since its variance is constant over the $x$-domain. 

Thus, the prior term $\mathcal{H}(s)$ is defined as a quadratic form with a stationary covariance matrix as in Eq. \eqref{eq: Stationary covariance matrix}. Rather than storing a dense matrix, we apply the Wiener-Khinchin theorem, stating that such a covariance matrix becomes diagonal in Fourier space\footnote{In our notation, we use $\tilde{f}(k)=\int \mathrm{d}x\:e^{ikx}f(x)$ for the forward and $f(x)=\int \mathrm{dk}/(2\pi)\: e^{-ikx}\tilde{f}(k)$ for the backward Fourier transform, where $k$ is the wavevector associated with the real space variable $x=-\mathrm{ln}(a)$.}:
\begin{equation}
\tilde{S}_{kq}^\mathrm{dev}=2\pi \cdot  \delta(k-q) p_s(k),
\end{equation}
where the diagonal elements are given by the signal power spectrum  $p_s(k)$. 

Following the argumentation of \cite{Porqueres_2017}, the a-priori signal power spectrum $p_s(k)$ is given by
\begin{equation} \label{eq: p_s(k) k-4 dependance} 
p_s(k)= v k^{-4},
\end{equation}
where $v$ determines the amplitude of field fluctuations. The exponent of $-4$ stems from an analysis of the total curvature of the field,
\begin{align}
\mathcal{C}:=\int \mathrm{d}x \: \vert \Delta  s(x)\vert ^2,   
\end{align}
which can be rewritten as $\mathcal{C}=s^{\dagger}(\Delta^{\dagger}\Delta) s$ by taking the scalar product to be the $L^2$-norm. Identifying the bracketed term as a covariance kernel via $S^{-1}:= v^{-1} \Delta^{\dagger}\Delta$ and Fourier-transforming leads to the exponent $-4$ in Eq. \eqref{eq: p_s(k) k-4 dependance}.

Thus, $S^{-1}$ takes on the function of a smoothing kernel, and the constant $v$ is a scaling constant that determines how much the probability of highly curved signal realizations is weighed down in the inference. This constant was set to $v=1$ in Eq. \eqref{eq: p_s(k) k-4 dependance} in \cite{Porqueres_2017}. Here, while we assume the same power-law dependence, we directly infer the smoothing constant $v$ from the data. More specifically, we replace the fixed parameter $v$ with a related quantity, the fluctuation parameter $b$, which models the amplitude of fluctuations around the mean linear trend of $s(x)$ (see Sect. \ref{subsection: Generative model prior}). In contrast to \cite{Porqueres_2017}, where the field variance at each point scales with $v$ and an integral contribution,  $\langle (s^x)^2\rangle_{(s)} = v\int_{k>0}k^{-4}\mathrm{d}k$, our model directly sets the variance as $\langle (s^x)^2\rangle_{(s)} = b^2$. 

Given the data, prior and likelihood terms, we only need to apply Bayes' Theorem to obtain the posterior distribution of the signal. However, analytical solutions are often not possible and, in many cases, the partition function $\mathcal{Z}$ is intractable analytically or computationally prohibitive \citep{knollmueller2018encoding}. 

Variational inference (VI) addresses this problem. The posterior distribution $\mathcal{P}\giventhat{s}{d}$ is approximated through a trial distribution using the Kullback-Leibler (KL) divergence to quantify the dissimilarity between the trial and posterior distribution. 

\cite{knollmueller2020metricgaussianvariationalinference} use a Gaussian for the trial distribution. However, the posterior is often non-Gaussian, leading to a sub-optimal approximation. In this work, we employ the \texttt{geoVI} algorithm \citep{Frank_2021}, which finds a coordinate transformation that effectively absorbs potential curvature of the posterior and finds a new coordinate system where the posterior is well approximated by a Gaussian distribution (see details in App. \ref{app: Working principle of geoVI}).

\subsection{Generative model prior}\label{subsection: Generative model prior}

As prior information, we assume that the universe evolves monotonically (expanding, contracting, or static) in a smooth fashion. The monotonicity of the evolution is captured through a linear model:
\begin{equation}
    s_{\mathrm{lin}} = \alpha x + \beta,
\end{equation}
where $\alpha$ and $\beta$ are random variables drawn from a Gaussian prior. Deviations from this linear evolution are modeled via a correlated field model (CFM). The CFM, introduced in \cite{arras_discrete_correlated_field} and \cite{Arras_2022}, generates field realizations from a parametric power spectrum with contributions from stochastic processes:
\begin{equation}
    s_{\mathrm{CFM}} = A\xi_s,
\end{equation}
where $A$ is the so-called amplitude operator and $\xi_s$ is a Gaussian standard variable, $\xi_s\hookleftarrow \mathcal{G}(0,\mathbf{1})$. In Fourier space, the diagonal elements of the amplitude operator are given by the amplitude spectrum $p_A(k)$, which is related to the signal power spectrum via $p_A(k)=\sqrt{p_s(k)}$. 

As discussed, we will assume that the amplitude spectrum follows a power-law: 
\begin{equation}
    p_A(k) = b\cdot \frac{k^{n}}{\sqrt{U}},
\end{equation}
where $U=\int_{k\neq0}\mathrm{d}k\hspace{1mm} k^{2n}$ normalizes the amplitude spectrum such that the real-space variance of the field is given by the fluctuations parameter $b$: $\langle (s^x)^2 \rangle = \int_{k\neq 0}\mathrm{d}k\hspace{1mm} p_A(k)^2 = b^2$. 

Adding the contribution by the linear model and the CFM gives us a signal realization:
\begin{equation}\label{eq: Generative Prior Model}
    s=s_{\text{lin}}+s_{\text{CFM}} =  \alpha x+\beta + A\xi_s.
\end{equation}
A flowchart of this forward model can be seen in Fig. \ref{fig: Computational graph of Correlated Field Model}.

\begin{figure}[ht]
    \centering
    \includegraphics[width=0.9\linewidth]{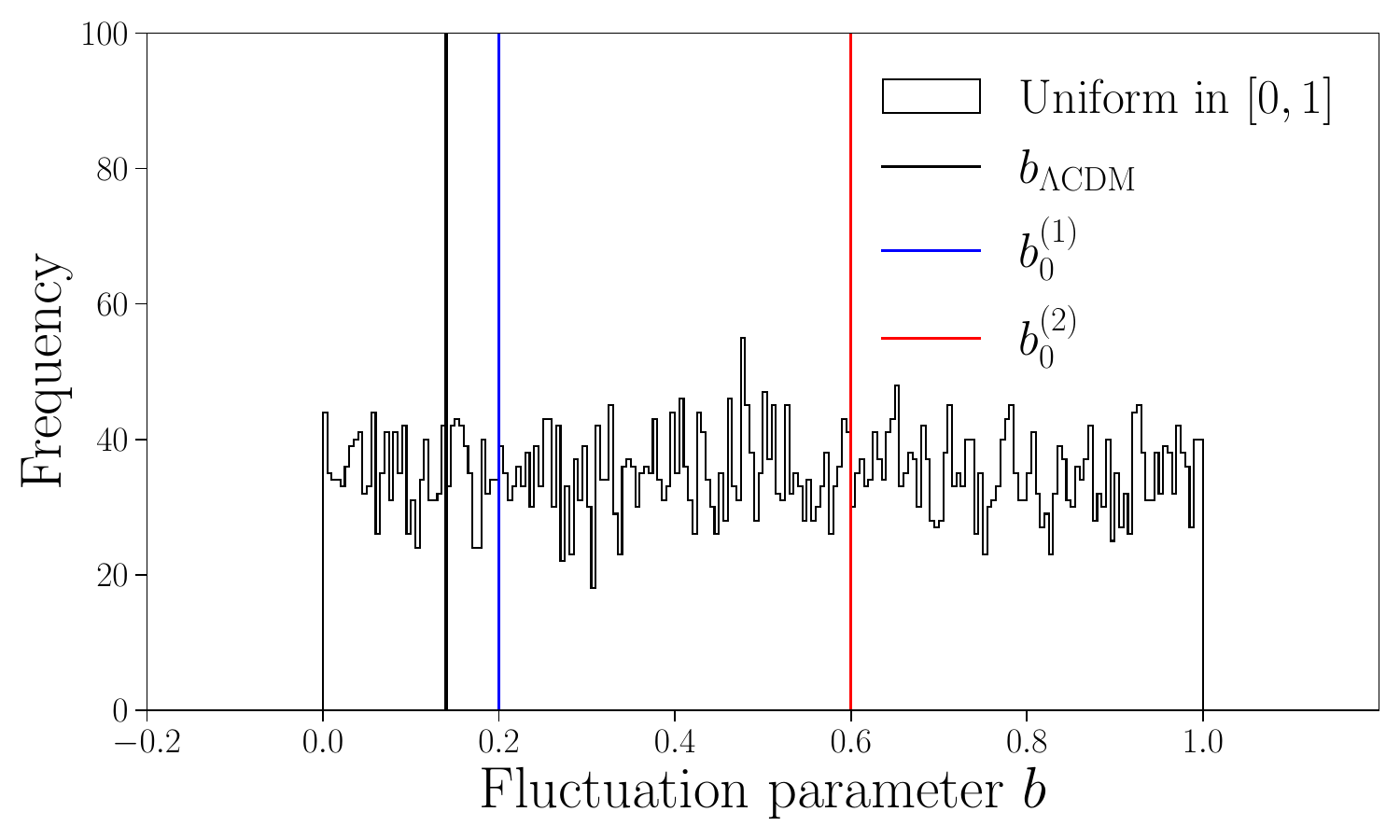}
    \caption{Prior samples for the fluctuations parameter, $b$, which are drawn from a uniform distribution in the interval $[0, 1]$. This parameter sets the a-priori expected level of field fluctuations around the mean field. We use the fluctuations parameter expected from a flat $\Lambda$CDM model, $b_{{\Lambda}\mathrm{CDM}}$ (black vertical line), to set an appropriate scale for this distribution heuristically (see App. \ref{sec: app: Construction of reference fields}). The blue and red vertical lines indicate two possible starting positions for the inference scheme, which were used in the real data reconstructions of this work.}
    \label{fig: Fluctuations histogram}
\end{figure}

Table \ref{table: Hyperparameters chosen for the generative prior model} shows the names, distributions, and prior mean and standard deviation of the hyperparameters. Here we discuss these prior choices: 
\begin{itemize}
    \item $\alpha$: the prior mean is heuristically set to $2$, encoding that it is more likely that the universe is expanding rather than contracting. The standard deviation is set to the same order of magnitude to ensure sufficient flexibility
    \item $\beta$: the prior mean is chosen as the $s(x=0)$ value of flat $\Lambda$CDM cosmology fields (cf. Fig. \ref{fig: Comparison Fields for flat LCDM}), again with a standard deviation in the same order of magnitude. This part of the prior is therefore informed by flat $\Lambda\mathrm{CDM}$ cosmology. However, we note that $\beta$ controls only the offset of the signal $s(x)$ not its shape and as discussed in the following sections, the offset cannot be faithfully recovered from SNIa alone
    \item $n$: is set to $-2$ according to Eq. \eqref{eq: p_s(k) k-4 dependance}. Notice that this is a factor of $1/2$ smaller than the exponent found in Eq. \eqref{eq: p_s(k) k-4 dependance} to account for the fact that $p_s(k)=\sqrt{p_A(k)}$
    \item $b$: drawn from a uniform distribution $\mathcal{U}$ in the interval $[0,1]$ to be maximally agnostic (see Fig. \ref{fig: Fluctuations histogram}). The order of magnitude of this parameter is obtained by comparing to a reference value, $b_{\mathrm{\Lambda CDM}}$, the expected fluctuations in the case of a flat $\mathrm{\Lambda CDM}$ cosmology. This reference value was obtained through a linear best-fit to a reference field constructed using the Planck2018 \citep{Planck_2018} cosmology parameters and found to be $b_{\mathrm{\Lambda CDM}}\approx 0.14$. Therefore, drawing $b$ from $\mathcal{U}(0,1)$ should allow for enough flexibility to reconstruct any non-$\Lambda$CDM features if preferred by the data. 
\end{itemize}

The CFM enforces periodic boundary conditions via Fast Fourier transforms of the power spectrum. To avoid artefacts due to periodic boundary conditions, we double the signal domain and introduce a zero-padding mask in the margins \citep{arras_discrete_correlated_field}. 

\begin{table}[ht]
    \centering
    \setlength{\tabcolsep}{12pt}
    \caption{Hyperparameters chosen for the generative prior model. Gaussian and uniform distributions are denoted by $\mathcal{G}(\mu^{\prime}, \sigma^{\prime})$ and $\mathcal{U}(l^{\prime}, u^{\prime})$, where $\mu^{\prime}$, $\sigma^{\prime}$, $l^{\prime}$ and $u^{\prime}$ are the mean, standard deviation, lower and upper bound respectively.}
    \label{table: Hyperparameters chosen for the generative prior model}
    \begin{tabularx}{0.5\textwidth}{X c l}  
        \toprule
        \textbf{Description} & \textbf{Symbol} & \textbf{Distribution} \\
        \midrule
        Slope of line model & $\alpha$ & $\mathcal{G}(2, 5)$ \\
        Offset of line model & $\beta$ & $\mathcal{G}(30,10)$ \\
        Power law exponent of amplitude spectrum & $n$ & $\mathcal{G}(-2, 10^{-16})$ \\
        Fluctuations & $b$ & $\mathcal{U}(0,1)$ \\
        \bottomrule
    \end{tabularx}
\end{table}

\section{Data}\label{sec: Data}

We use our inference method to analyze three main SN Ia data sets: Union2.1 \citep{Suzuki_2012}, Pantheon+ \citep{Pantheon_2022} and DESY5 \citep[Dark Energy Survey Year 5,][]{DES2024, DESY5}. We further discuss the impact of a recent recalibration of DESY5, labeled "DESY5-Dovekie" \citep{DES_Dovekie}, on our reconstruction results. 

The data sets consist of distance moduli $\mu$ obtained by fitting the SN light curves with spectral energy distribution templates \citep{SALT3_Kenworthy_2021},
\begin{equation}
    \mu = m+\chi  x_1 - \psi y -M-\delta_{\mu-\mathrm{bias}},
\end{equation}
where $m=-2.5\mathrm{log}_{10}(x_0)$ is the apparent magnitude and $M$ is the absolute magnitude; $x_0$, $x_1$ and $y$ parametrize the lightcurve; and $\chi$, $\psi$ and $\delta_{\mu-\mathrm{bias}}$ are calibration parameters. 

For the distance modulus, the absolute magnitude $M$ is degenerate with $H_0$. Therefore, many analyses combine the two parameters as  (e.g., \cite{Perlmutter_1999}, \cite{rubin2023unionunitycosmology2000} and \cite{DESY5}),
\begin{equation}\label{eq: Magnitude Hubble degeneracy}
    \mathcal{M}=M+5\mathrm{log}_{10}(c/H_0),
\end{equation}
which is jointly inferred with the cosmological parameters. In this case, the distance moduli can be obtained only by assuming a value for $H_0$. 

Union2.1
and 
DESY5
assume a value of $h := H_0/(\mathrm{100\: km/s/Mpc})=0.7$, while 
Pantheon+
uses Cepheids as a primary distance anchor to infer a value of $H_0$ from low-redshift measurements, allowing to lift the degeneracy in Eq. \eqref{eq: Magnitude Hubble degeneracy}. 

\begin{figure}[ht]
    \centering
    \includegraphics[width=0.9\linewidth]{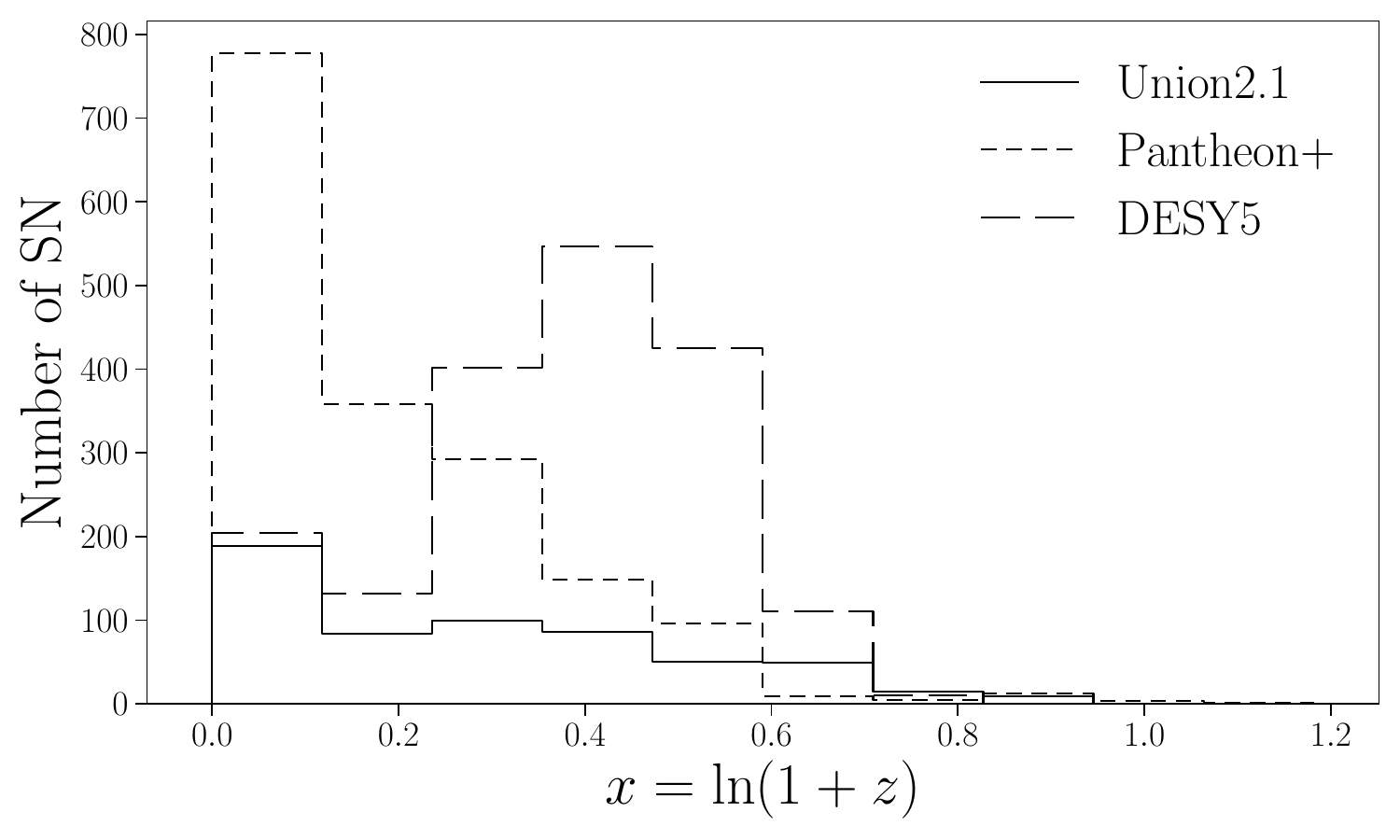}
    \caption{Histograms of the datasets binned in ten $x$ bins. While the Pantheon+ sample dominates the low redshift regime, the new DESY5 sample shows a significant boost in the number of mid-to-high redshift SN samples. The DESY5-Dovekie histogram (not shown) deviates only marginally from the original DESY5 version.}
    \label{fig: Hubble Diagram}
\end{figure}

Figure \ref{fig: Hubble Diagram} shows the redshift distributions of the main data sets, and Table  \ref{table: Redshift range of used datasets} indicates their redshift range.

\begin{table}[ht]
    \centering
    \setlength{\tabcolsep}{12pt}
    \caption{Redshift range and number of datapoints of the data sets analyzed in this work.}
    \label{table: Redshift range of used datasets}
    \begin{tabularx}{0.5\textwidth}{X c c}  
        \toprule
        \textbf{Data set} & $\mathbf{(z_{\mathrm{min}}, z_{\mathrm{max}}, \overline{z})}$ & \textbf{\# Datapoints} \\
        \midrule
        Union2.1 & $(0.015, 1.414, 0.36)$ & 580 \\
        Pantheon+ & $(0.00122, 2.26, 0.22)$ & 1701 \\
        DESY5 & $(0.025, 1.12, 0.46)$ & 1829 \\
        DESY5-Dovekie & $(0.025, 1.14, 0.46)$ & 1820\\
        \bottomrule
    \end{tabularx}
\end{table}

\section{Results}\label{sec: Results}

In this Section, we apply the \texttt{charm2} algorithm to synthetic and real data, to validate our inference method and reconstruct the posterior mean field $s=\mathrm{ln}(\rho/\rho_0)$ from observations. Using the evidence lower bound (ELBO) measure, often encountered in variational approaches, we determine whether there exists a preference for non-$\Lambda\mathrm{CDM}$ features in any data sets analyzed here. Finally, we discuss the effect of the recalibrated DESY5-Dovekie sample on our reconstructions.

\subsection{\texorpdfstring{Impact of $b$ initialization on field reconstruction}{Impact of b initialization on field reconstruction}}\label{subsec: Impact of b initialization on field reconstruction}

To improve the convergence of the algorithm, a reasonable initial parameter vector must be chosen. We find that, in practice, the reconstructed posterior mean field shows some dependence on the choice of the initial fluctuation parameter $b_0$. Yet, reconstructions with different initial $b_0$ values generally agree within their reconstruction uncertainties. 

Since the data values are line-of-sight integrals over $x=-\mathrm{ln}(a)$, signal fields that oscillate at a high rate around their mean line give approximately the same data as signals with damped oscillations. Therefore, SNe alone are a suboptimal probe to constrain the field fluctuations of $s=\mathrm{ln}(\rho/\rho_0)$.

\begin{figure}[ht]
    \centering
    \includegraphics[width=0.9\linewidth]{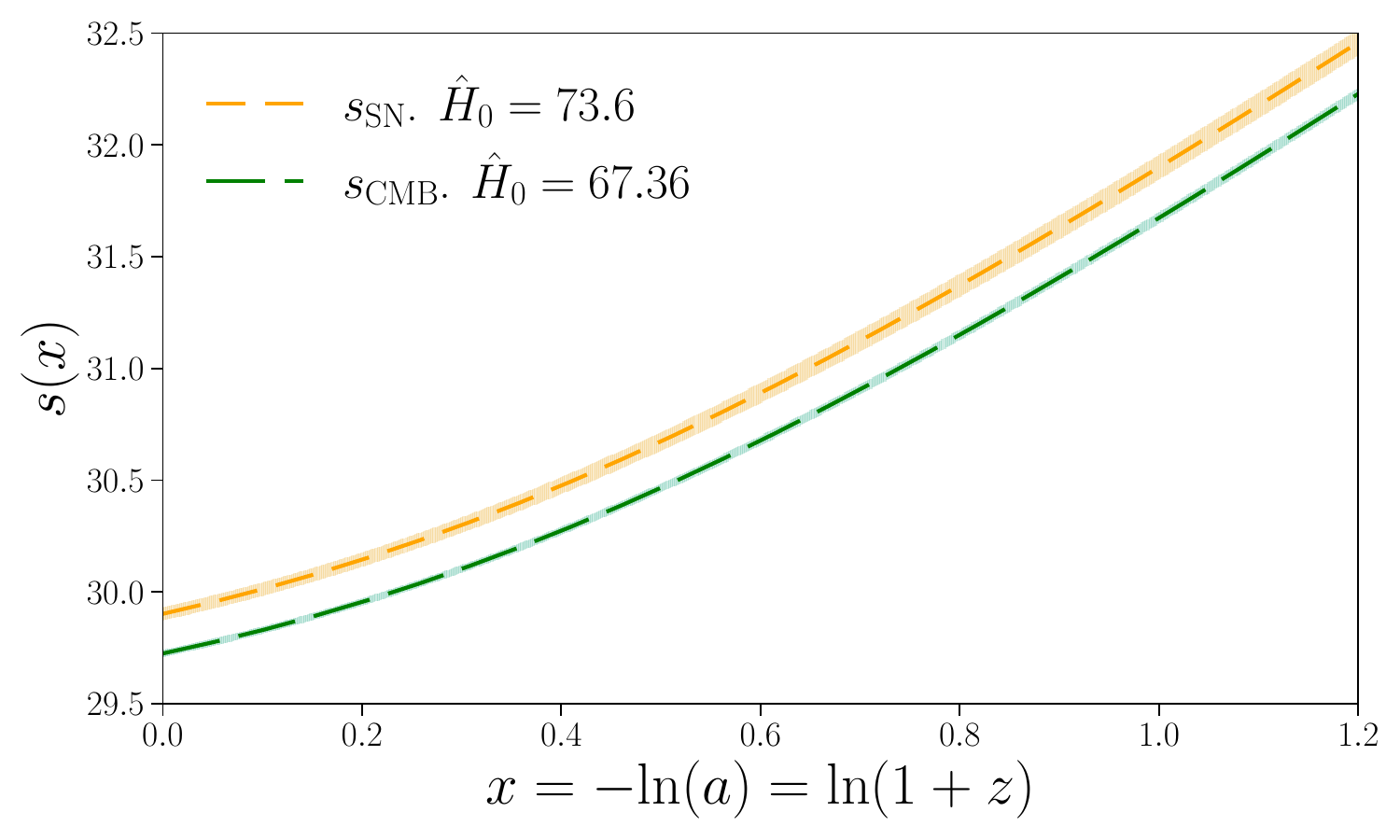}
    \caption{Mean field and estimated $1\sigma$ region for flat $\Lambda$CDM models evaluated with the $\Omega_m$ and $H_0$ values from Planck2018 (green) and Pantheon+SH0ES (orange). 
    The $1\sigma$ contour was estimated through Gaussian error propagation of the uncertainties of the values in Table \ref{table: Adopted values of H0 and Omega_m for comparison of reconstructions to flat LCDM model} into Eq. \protect\eqref{eq: Translation of flat lcdm into signal field}, which assumes $H_0$ and $\Omega_m$ are independent. These fields serve as benchmarks for our reconstructions.}
    \label{fig: Comparison Fields for flat LCDM}
\end{figure}

We thus present our results in terms of the initial fluctuation parameter $b_0$, showing reconstructions where it was initialized at lower values,  $b_0^{(1)}=0.2$ and higher values, $b_0^{(2)}=0.6$ (see vertical colored lines in Fig. \ref{fig: Fluctuations histogram}). A first estimate of the field fluctuations can be obtained by fitting a linear model to the benchmark $\Lambda\mathrm{CDM}$ curves seen in Fig. \ref{fig: Comparison Fields for flat LCDM}. This provides a value for the residual variance, $b_{\Lambda\mathrm{CDM}}=0.14$, on which we base our suggestion for the initial values: $b_0^{(1)}=0.2$ lies close to the theoretical $b_{\Lambda\mathrm{CDM}}$ value while providing some additional flexibility, whereas $b_0^{(2)}=0.6$ represents a significant potential deviation from flat $\mathrm{\Lambda\mathrm{CDM}}$. 





\subsection{Synthetic reconstructions}\label{subsec: Synthetic reconstruction}

\begin{figure}[ht]
    \centering
    \includegraphics[width=1\linewidth]{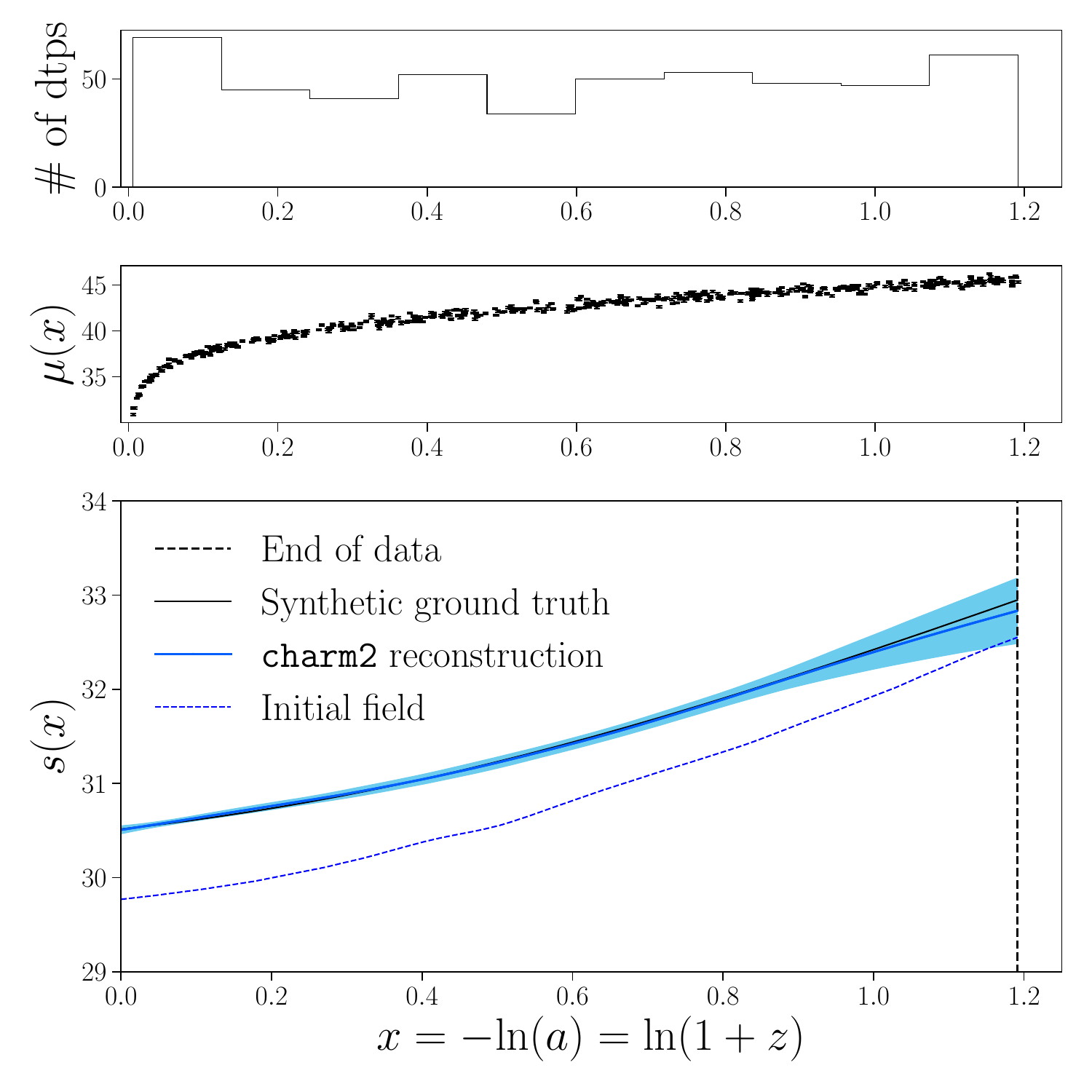}
    \caption{Reconstruction from a synthetic data set with $\texttt{charm2}$. From top to bottom: Histogram of the number of data points; the data plotted as a function of the $x$-coordinate; and the initial field (dashed blue line) along with the posterior mean field (solid blue line) with its $1\sigma$ uncertainty band (shaded blue region) and the ground truth (solid black line). The inference was initialized with a fluctuation parameter of $b_0=0.2$. The noise variance in the simulated data is set to $\sigma_n^2=0.1$, with the errorbars vanishingly small compared to the scale of the $y$-axis.}
    \label{fig: Synthetic Reconstruction uniform data b0 = 0.2}
\end{figure}

We validate the inference method with synthetic data sets generated through a flat $\Lambda$CDM model with $\Omega_\mathrm{m0}=0.3$. Each data set contains $500$ datapoints, but the distribution over redshift varies as:

\begin{enumerate}
    \item Data set 1: the redshifts are sampled from a uniform distribution within the redshift range, $\mathcal{P}(z) = \mathrm{cst.}$
    \item Data set 2: the density of data points decays with redshift exponentially. The redshifts are sampled from the normalized distribution $\mathcal{P}(z) = \frac{1}{2}e^{-\frac{1}{2}z}$ and rescaled to the corresponding redshift range.
\end{enumerate}

The redshifts and signal fields are then mapped into the data space via the response operator, Eq. \eqref{eq: Instrument Response Operator}, and Gaussian noise is added to mimic the data dispersion.

The noise variance was set to $0.1$, motivated by the mean variance of data points in the Union2.1 compilation ($\approx$ 0.07).

\begin{figure*}[htbp]
  \centering
  \subfloat[Initial fluctuations parameter $b_0=0.2$.]{
    \includegraphics[width=0.48\textwidth]{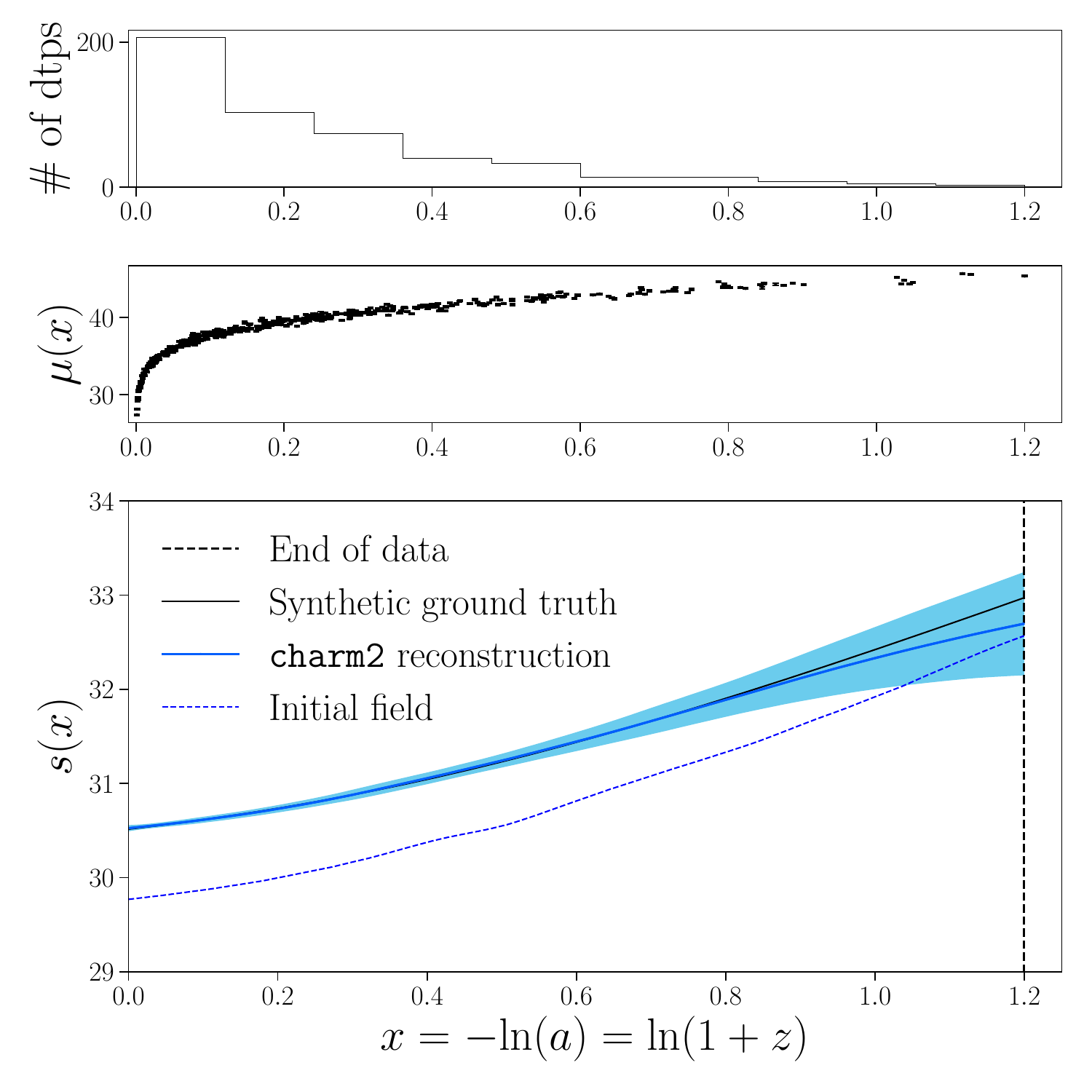}
    \label{subfig: Synthetic Reconstruction exponential data b0=0.2}
  }
  \hfill
  \subfloat[Initial fluctuations parameter $b_0=0.6$.]{
    \includegraphics[width=0.48\textwidth]{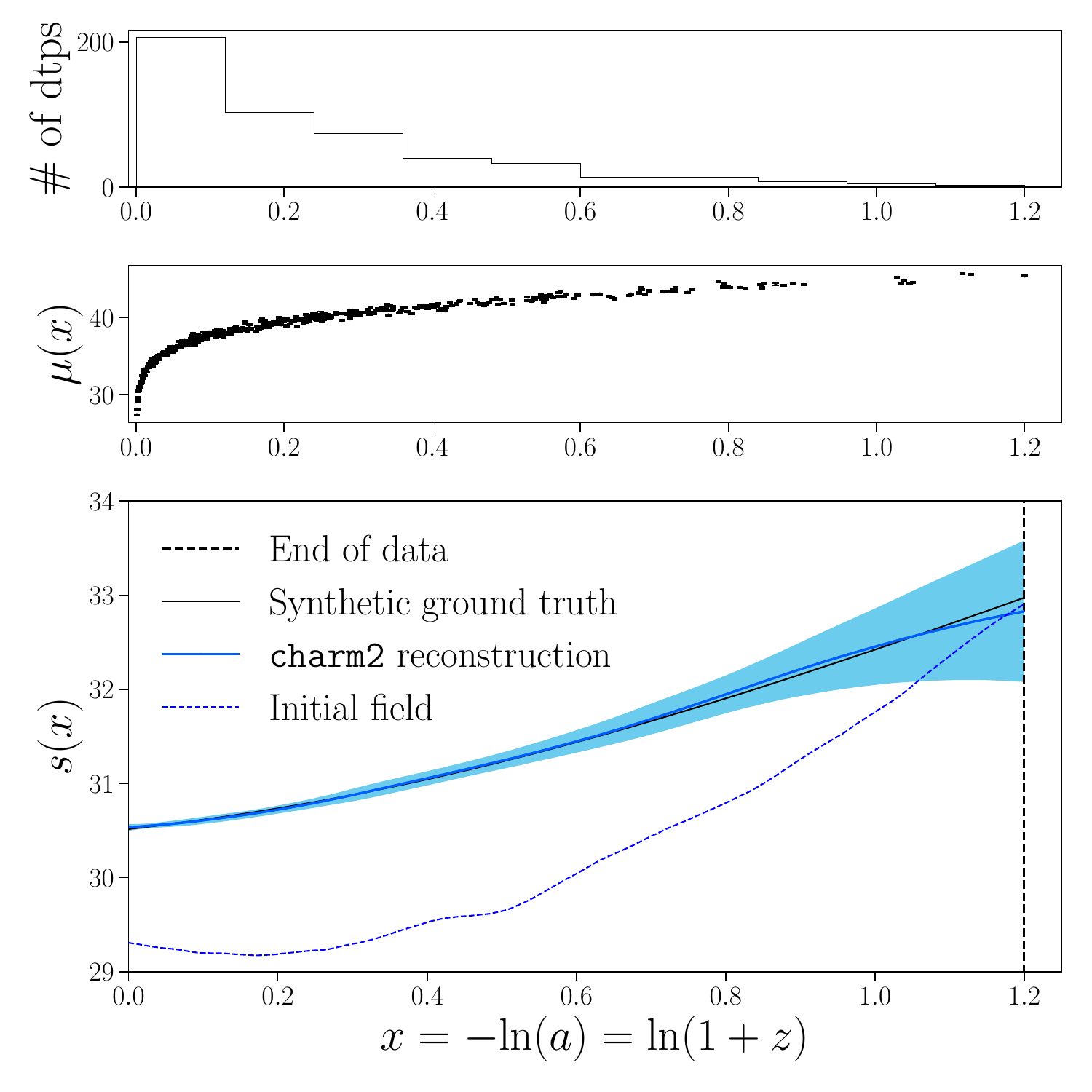}
    \label{subfig: Synthetic Reconstruction exponential data b0=0.6}
  }
  \caption{Same as Fig.~\ref{fig: Synthetic Reconstruction uniform data b0 = 0.2} with an exponentially falling redshift distribution and with an initial fluctuation parameter of $b_0=0.2$ (left) and $b_0=0.6$ (right).}
  \label{fig:my_two_plots}
\end{figure*}

For an initial fluctuation parameter of $b_0=0.2$, as shown in Fig. \ref{fig: Synthetic Reconstruction uniform data b0 = 0.2} (data set 1) and \ref{subfig: Synthetic Reconstruction exponential data b0=0.2} (data set 2), our inference method recovers the true signal across the redshift range for both synthetic data sets. For data set 2 (Fig. \ref{subfig: Synthetic Reconstruction exponential data b0=0.2}), the reconstruction uncertainty increases at high $x$, where fewer data points are available.

By comparing Fig. \ref{subfig: Synthetic Reconstruction exponential data b0=0.2} ($b_0=0.2$) and \ref{subfig: Synthetic Reconstruction exponential data b0=0.6} ($b_0=0.6$), we can assess that the analysis of synthetic data is robust against different initializations of the fluctuation parameter, as the posterior mean fields agree within their reconstruction uncertainty.

\subsection{Union2.1, Pantheon+ and DESY5 reconstructions} 
\begin{figure*}[ht]
    \centering
    \includegraphics[width=1\linewidth]{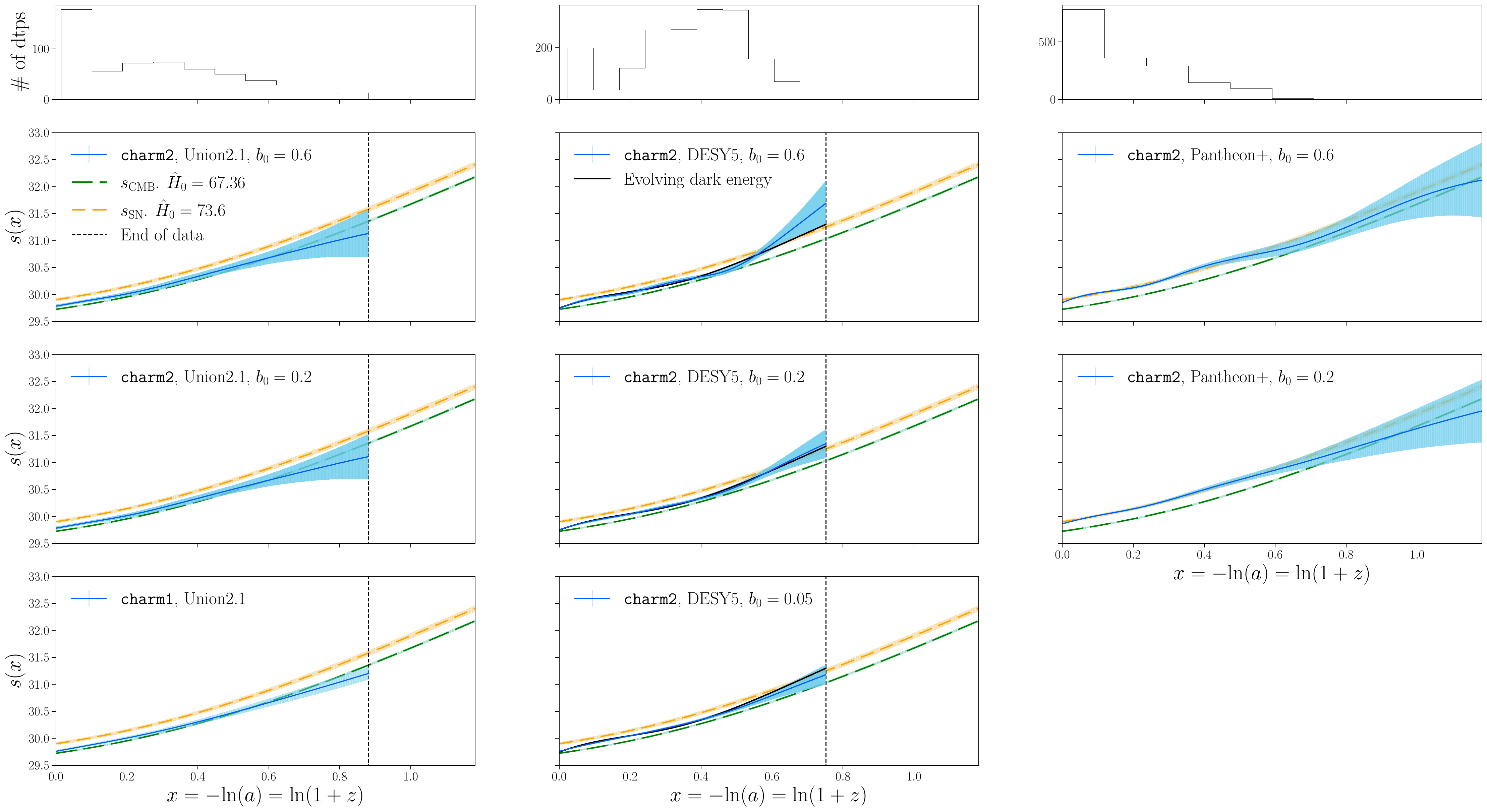}
    \caption{\texttt{charm2} reconstructions for 
    Union2.1
    (left column), 
    DESY5
    (middle column) and 
    Pantheon+
    (right column) and for different initial values of the fluctuation parameter $b_0$ (rows). In each panel, the posterior mean-field (solid blue line) and its $1\sigma$ confidence region (shaded blue region) are displayed alongside the reference fields of Fig. \protect\ref{fig: Comparison Fields for flat LCDM} (dashed orange and green lines). Additionally, for the 
    DESY5
    reconstruction, the signal field corresponding to the evolving dark energy model suggested by \cite{DESY5} ($(\Omega_m, w_0, w_a)=(0.495, -0.36, -8.8)$) is shown as the black solid line. A comparison with the previous  \texttt{charm1} analysis of Union2.1 is shown in the left column (lower panel), where the signal definition is adapted as described in App. \protect\ref{app: Note on converting charm1 fields}.
    }\label{fig: Final Reconstructions}
\end{figure*}

We applied \texttt{charm2} to the 
Union2.1,
Pantheon+
and 
DESY5
data sets. We compare our reconstructions to flat $\Lambda$CDM reference fields shown in Fig. \ref{fig: Comparison Fields for flat LCDM}, $s_{\mathrm{CMB}}$ and $s_{\mathrm{SN}}$, corresponding to the signal for the cosmological parameters found by the Planck2018 \citep{Planck_2018} and the Pantheon+SH0ES \citep{Pantheon_Cosmological_Analysis} analyses, respectively (see App. \ref{sec: app: Construction of reference fields} for construction).

The reconstructed fields for the three data sets are presented in Fig. \ref{fig: Final Reconstructions}, alongside histograms of their respective redshift distributions.

\needspace{2\baselineskip} 
\vspace{0.5em}
\textbf{\underline{Union2.1}}
\vspace{0.5em}

In the first data set to be analyzed, we compare $\texttt{charm2}$ to its predecessor code, $\texttt{charm1}$, which was only applied to Union2.1 data, as this was the most up-to-date SNIa sample at that time. While the reconstructed fields are compatible, \texttt{charm2} provides more realistic uncertainties by better accounting for the nonlinearities of the response operator, which were neglected in \cite{Porqueres_2017}. As a result, we see that the reconstruction uncertainty at high redshifts was significantly underestimated in the $\texttt{charm1}$ reconstruction, demonstrating the advantage of non-Gaussian posterior approximation techniques. We also show the effect of changing \texttt{charm2}’s initial fluctuations parameter from $b_0 = 0.2$ to $b_0 = 0.6$. The results are almost identical, indicating that the $\texttt{charm2}$ analysis of the Union2.1 data set is robust to the choice of the proposed $b_0$ initial values.

Note that the Union2.1 data assumes a fiducial value of $H_0=70\: \mathrm{km/s/Mpc}$ to calibrate the distance modulus $\mu$ (see Sect. \ref{sec: Data}). This $H_0$ value affects the reconstruction at $z=0$, making it lie closer to the $s_{\mathrm{CMB}}$ field. Since $s_\mathrm{CMB}$ however assumes $H_0=67.37\:\mathrm{km/s/Mpc}$, the two curves are slightly offset from each other.

\needspace{2\baselineskip} 
\vspace{0.5em}
\textbf{\underline{Pantheon+}}
\vspace{0.5em}

The $\texttt{charm2}$ analysis of Pantheon+ data is also robust to the choice of initial $b_0$ value, providing reconstructed fields that are compatible within the $1\sigma$ uncertainty, with slightly larger fluctuations around the mean line in the $b_0=0.6$ reconstruction. Further, the Pantheon+ reconstruction is compatible with the $s_{\mathrm{SN}}$ field, but deviates from $s_{\mathrm{CMB}}$ by more than $2\sigma$ at $x<0.6$, a manifestation of the Hubble Tension, as discussed, e.g., in \cite{Riess_2022}.
 
\needspace{2\baselineskip} 
\vspace{0.5em}
\textbf{\underline{DESY5}}
\vspace{0.5em}

The final data set analyzed in this work is the DESY5 SNIa sample, which has been reported to slightly prefer an evolving dark energy (EDE) component that follows an equation of state
\begin{equation}
    w_{\Lambda}(a)=w_0+w_a(1-a)
\end{equation}
where $w_0$ sets the baseline and $w_a$ controls the time evolution \citep{DESY5}. Using this model, often referred to as $w_0w_a\mathrm{CDM}$, \cite{DESY5} found
\begin{equation}
    (\Omega_m, w_0, w_a)\approx (0.495, -0.36, -8.8)
\end{equation}
for the mean best-fit parameters. Using a fiducial Hubble constant of $H_0=68.1\:\mathrm{km/s/Mpc}$\footnote{Through a variational inference $w_0w_a\mathrm{CDM}$ fit to the DESY5 data, we found a posterior value of $\hat{H}_0=\: 68.1\pm 0.4$ and thus heuristically set $\hat{H}_0\approx 68.1$}, we calculate the parametric signal field expected from this EDE scenario, $s_{\mathrm{EDE}}$, and use it as an additional benchmark for the DESY5 reconstructions.

The $\texttt{charm2}$ reconstruction of the original DESY5 data shows a feature at $x\sim [0.45,0.75]$ (corresponding to redshift $z\sim [0.57,1.17]$) that is in tension with both the $s_{\mathrm{SN}}$ and the $s_{\mathrm{CMB}}$ flat $\mathrm{\Lambda CDM}$ fields, while in agreement with the evolving dark energy $s_\mathrm{EDE}$ field for both the $b_0=0.2$ and $b_0=0.6$ initial conditions, as well as a very low initial fluctuation value of $b_0=0.05$. The influence of the initial $b_0$ value on the posterior mean field is most clearly seen in the analysis of the DESY5 data. Note, however, that a value of $b_0=0.05$ leads to posterior samples that only barely fluctuate around their mean linear trend, which in effect corresponds to a linear fit of the expansion history and is therefore suboptimal from a theoretical point of view. Therefore, higher values such as $b_0=0.2$ and $b_0=0.6$, should be considered more reliable.

\begin{figure}[htbp]
    \centering
    \includegraphics[width=1\linewidth]{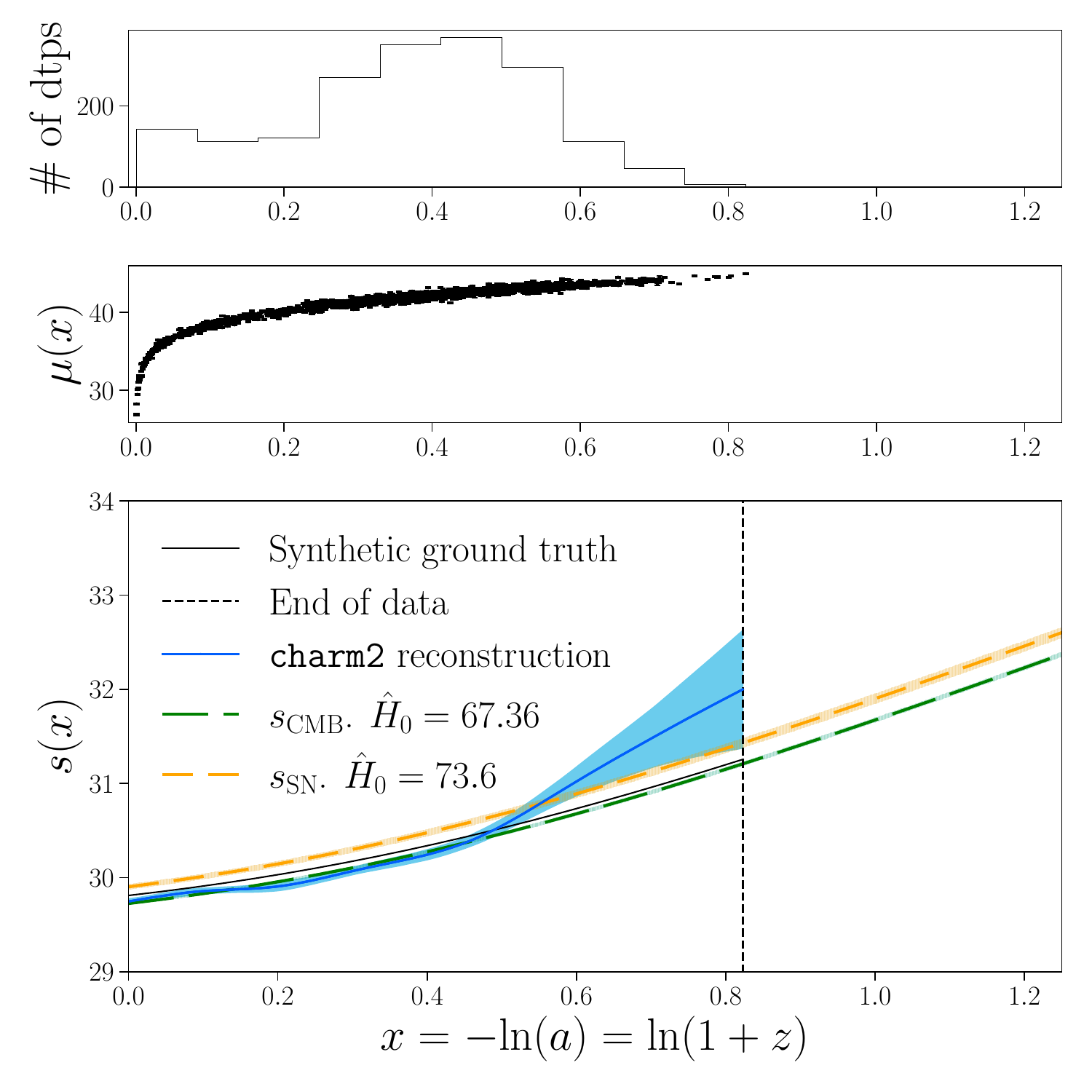}
    \caption{A $b_0=0.6$ $\texttt{charm2}$ reconstruction for a synthetic data set generated with a flat $\Lambda$CDM model with $\Omega_{\mathrm{m0}}=0.3$  and $H_0=70.3\:\mathrm{km/s/Mpc}$, where all data points $x<0.6$ were systematically shifted by $+0.1\hspace{1mm}\mathrm{mag}$. }
    \label{fig: EDE as a bump in the data}
\end{figure}

We tested whether an evolving-dark-energy-like feature could be caused by a systematic offset in magnitude between low- and high-redshift SNe in the sample, even though the ground truth is flat $\Lambda\mathrm{CDM}$. This is motivated by the systematic offset of $0.04\hspace{1mm}\mathrm{mag}$ pointed out by \cite{Efstathiou2024} between the low-and high-redshift data of the 
Pantheon+
and 
DESY5
compilations. For the necessary mock study, we generated data according to a flat $\Lambda$CDM model and introduced a shift of $+0.1\hspace{1mm}\mathrm{mag}$ for all data points falling under an arbitrary threshold $x < 0.6$. Figure \ref{fig: EDE as a bump in the data} shows that in such a case, the reconstructed mean field exhibits a deviation visually similar to the one observed in our DESY5 reconstructions at high redshifts, indicating that systematics might cause the observed feature. However, we could not identify a systematic offset at low redshift $x<0.1$ that reproduces such a feature, which is the regime where \cite{Efstathiou2024} found  $\langle m_{\text{Pantheon+}} - m_{\text{DESY5}} \rangle = (-0.051 \pm 0.007)\hspace{1mm}\mathrm{mag}$. 
Furthermore, \cite{MariaVincenzi2025Comparingdessn5yrpantheonsn} clarified that the discrepancy between DESY5 and Pantheon+ arises from changes in the DESY5 analysis pipeline, leaving a potential systematic offset of the order of $0.008\hspace{1mm}\mathrm{mag}$. Our tests showed that such a systematic offset would no longer have a significant impact on our reconstruction results, in which case the reconstructions would indeed indicate an evolving dark energy scenario.

\begin{figure}[ht]
    \centering
    \includegraphics[width=1\linewidth]{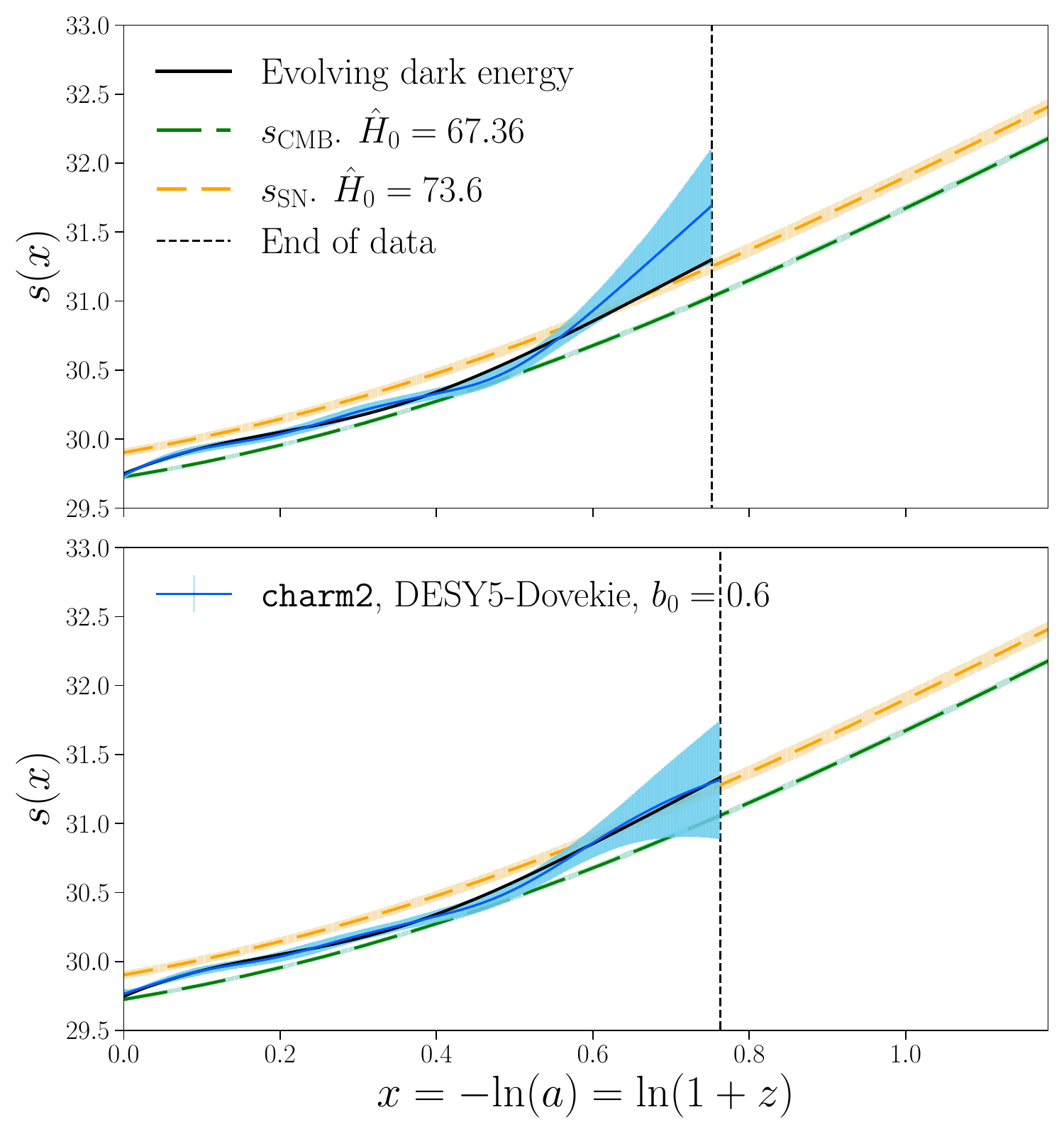}
    \caption{A comparison between the $b_0=0.6$ $\texttt{charm2}$ reconstructions for the original DESY5 and the updated DESY5-Dovekie data set.}
    \label{fig: Comparison of b0 0.6 des vincenzi vs des dovekie}
\end{figure}

Further, \cite{DES_Dovekie} recently introduced an updated version of DESY5, making improvements to photometric calibration, error handling, and the treatment of dust reddening. We reran the $\texttt{charm2}$ analysis on this updated data set and find that for $b_0=0.05$ and $b_0=0.2$ the DESY5-Dovekie reconstructions change only marginally compared to their DESY5 counterparts in Fig. \ref{fig: Final Reconstructions}. In the more flexible $b_0=0.6$ case, we find that the energy values of the posterior mean field are significantly reduced at high redshifts when going from DESY5 to DESY5-Dovekie, cf. Fig. \ref{fig: Comparison of b0 0.6 des vincenzi vs des dovekie}. This deexcitation of the posterior mean field of the nonparametric reconstruction can therefore be attributed to the corrected systematics in the DESY5-Dovekie catalog. However, it remains to be assessed whether the data genuinely prefer any nonparametric excitations of the signal field.

\subsection{Model selection}

The evidence term in Bayes' Theorem, 
\begin{equation}\label{eq: Definition of evidence via integration}
    \mathcal{P}(d)=\int \mathrm{D}s\:  \mathcal{P}(d\vert s) \mathcal{P}(s),
\end{equation}
where $\int\mathrm{D}s$ denotes an integral over all $s$ field configurations, gives the probability that the observed data was generated by the assumed prior model. It is thus a useful tool for selecting between two physical models that are consistent with the data, with the model producing the higher evidence term being preferred by the data. However, in our high-dimensional, nonparametric setting, the integration in Eq. \eqref{eq: Definition of evidence via integration} is computationally intractable and we instead use a proxy for the evidence known as the evidence lower bound, ELBO, which provides a lower bound on the log evidence:
\begin{equation}
    \mathrm{ln}\big(\mathcal{P}(d)\big)\geq \mathrm{ELBO},
\end{equation}
see App. \ref{app: Details on the evidence lower bound}.
These two quantities differ by the information difference in nits (natural information units) between the posterior and its variational approximation \citep{GuardianiCovid}. If the variational approximation captures the true posterior well, the ELBO is approximately equal to the log evidence. Given the complexity of distributions $\texttt{geoVI}$ can represent, we find the ELBO to be a very good proxy for the evidence.  

The averaging procedure involved in calculating the ELBO (see Eq. \eqref{eq: terms of ELBO}) naturally introduces a sampling uncertainty which we quote in Tab. \ref{table: ELBO of all cosmological reconstructions}, where the ELBOs computed from flat $\Lambda\mathrm{CDM}$, flat $w_0w_a\mathrm{CDM}$, and our nonparametric model fit are displayed for all data sets. Using the ELBO, we find no conclusive evidence for non-$\Lambda\mathrm{CDM}$ features in the SNIa samples analyzed in this work.

This indicates that flat $\Lambda\mathrm{CDM}$ can explain the data extremely well with only $2$ parameters ($H_0$ and $\Omega_m$). In contrast, the extra parameters added in flat $w_0w_a\mathrm{CDM}$ and our nonparametric model do not increase goodness of fit sufficiently to counterbalance the penalty by the Occam's Razor term encoded in the ELBO. 

\begin{figure}[ht]
    \centering
    \includegraphics[width=1\linewidth]{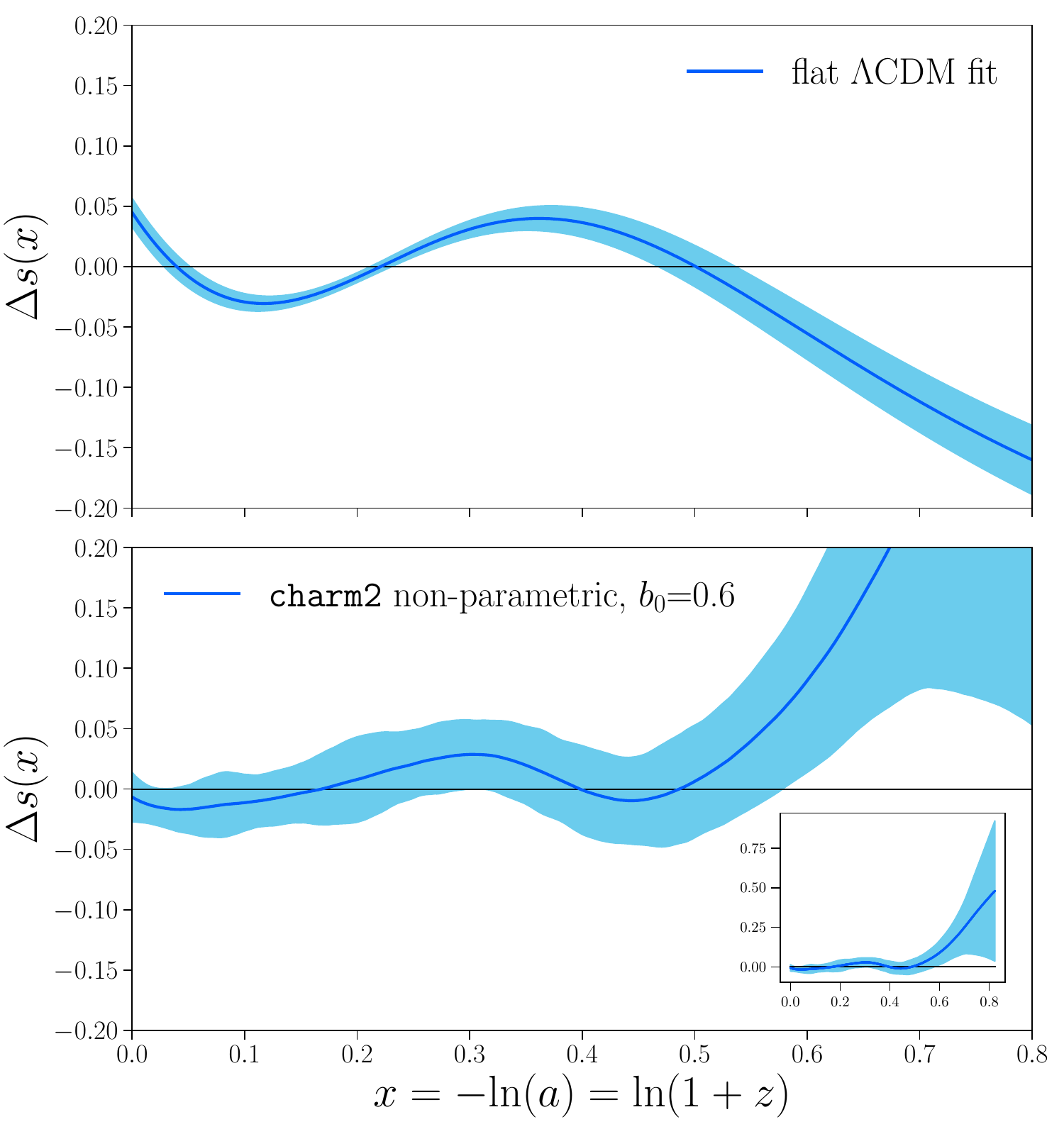}
    \caption{Residuals between a flat $\Lambda\mathrm{CDM}$ fit (upper panel) and a nonparametric $b_0=0.6$ $\texttt{charm}$ reconstruction (lower panel) and the ground truth, corresponding to an evolving dark energy model. The inset axis shows a zoomed out version of the nonparametric reconstruction. The synthetic data is generated with the real DESY5 survey noise covariance.}
    \label{fig: Evolving dark energy as ground truth: lcdm vs npa}
\end{figure}

\begin{figure}[htbp]
    \centering
    \includegraphics[width=1\linewidth]{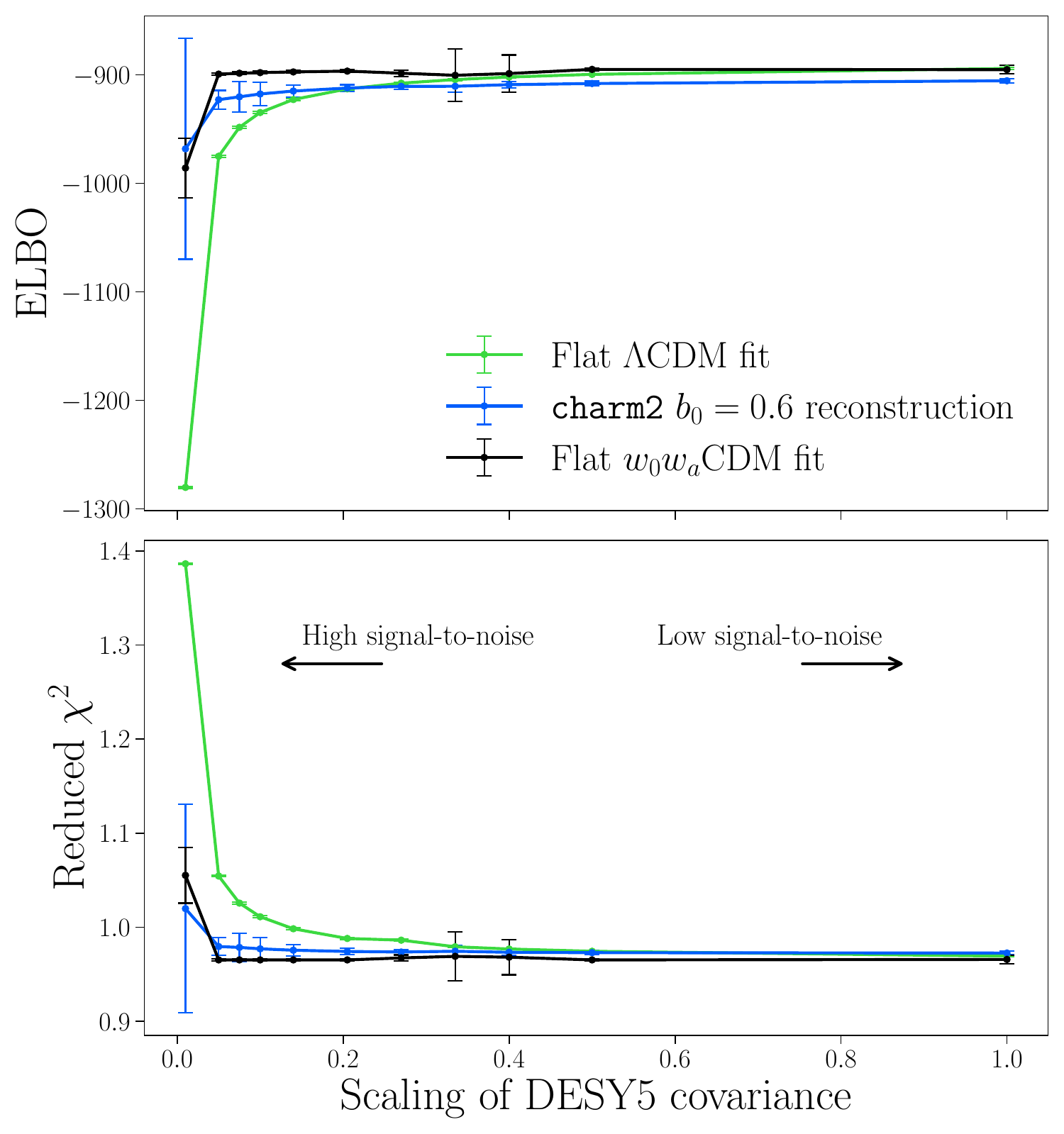}
    \caption{Evolution of the ELBO and reduced likelihood $\chi^2$ with varying noise level. The synthetic data was generated according to an evolving dark energy ground truth and the real DESY5 survey noise covariance matrix. For each point, the full noise covariance was multiplied by a constant factor, given by the corresponding $x$-axis value (high signal-to-noise on the left, low on the right).}
    \label{fig: ELBOs and chi2}
\end{figure}

However, the ELBO might be a suboptimal model selection statistic due to an insufficient signal-to-noise ratio. We studied a synthetic dataset generated with an evolving dark energy scenario and the DESY5 noise covariance matrix. The results are shown in Fig. \ref{fig: Evolving dark energy as ground truth: lcdm vs npa}. While the nonparametric reconstruction encloses $68\%$ of the ground truth within the $1\sigma$ uncertainty regions, the flat $\Lambda\mathrm{CDM}$ fit encloses only $13\%$, leading to a confident rejection of the evolving dark energy ground truth. The evidences in this scenario, 
\begin{align}
    \mathrm{ELBO}_{\text{flat }\Lambda\mathrm{CDM}} &= -894.19 \pm 0.85, \\ 
    \mathrm{ELBO}_{\text{non-param. }} &=  -905.5 \pm 1.8,
\end{align}
support this overconfident rejection of evolving dark energy. While including other cosmology probes might increase the ELBO reliability, at the current noise level of the DESY5 survey, we find that for SNIa alone the noise covariance would need to be smaller by a factor of $\gtrsim 7$ for the ELBO diagnostic to favor the nonparametric model which more accurately represents the ground truth over flat $\Lambda\mathrm{CDM}$ (see Fig. \ref{fig: ELBOs and chi2}).  If the ELBO is a sufficiently accurate proxy, this argument extends to the evidence itself, underscoring the importance of synthetic control studies for assessing its reliability in model selection.

To conclude, we find no data-driven evidence favoring nonparametric deviations from flat $\Lambda\mathrm{CDM}$ in the data sets analyzed in this work, which are the Union2.1, Pantheon+, DESY5 and DESY5-Dovekie supernova samples. At the same time, we caution that for SNIa alone, the evidence might be a sub-optimal model selection statistic for nonparametric models at current noise levels.

\begin{table*}[ht]
    \centering
    \setlength{\tabcolsep}{12pt}
    \caption{ELBO of all $\texttt{charm2}$ reconstructions shown in Fig. \protect\ref{fig: Final Reconstructions} compared against flat $\Lambda\mathrm{CDM}$ and flat $w_0w_a\mathrm{CDM}$ fits of the data. All errors were rounded to two significant figures.}
    \label{table: ELBO of all cosmological reconstructions}
    \begin{tabularx}{\textwidth}{X c c c c}
        \toprule
        & \textbf{Union2.1} & \textbf{DESY5} & \textbf{DESY5-Dovekie} & \textbf{Pantheon+} \\
        \midrule
        Flat $\Lambda\mathrm{CDM}$ 
        & $-278.9 \pm 1.2$ 
        & $-829.02 \pm 0.77$ 
        & $-824.8 \pm 1.3$ 
        & $-885.2 \pm 1.2$ \\

        Flat $w_0w_a\mathrm{CDM}$  
        & $-282.2 \pm 6.5$ 
        & $-829.0 \pm 1.2$ 
        & $-826.0 \pm 1.4$ 
        & $-886.0 \pm 1.3$ \\

        \midrule
        nonparametric $\texttt{charm2}$ \\

        \hspace{2mm}$b_0=0.6$  
        & $-284.3 \pm 1.3$  
        & $-835.0 \pm 2.6$  
        & $-832.1 \pm 3.9$  
        & $-889.3 \pm 3.1$ \\

        \hspace{2mm}$b_0=0.2$  
        & $-284.2 \pm 1.2$  
        & $-834.1 \pm 2.3$  
        & $-830.2 \pm 2.2$  
        & $-889.6 \pm 3.6$ \\

        \hspace{2mm}$b_0=0.05$  
        &  
        & $-834.6 \pm 2.5$  
        & $-831.0 \pm 2.3$  
        &  \\
        \bottomrule
    \end{tabularx}
\end{table*}
   
\section{Conclusions}\label{sec: Conclusions}

We have presented \texttt{charm2}, a Bayesian method to reconstruct the cosmic expansion from supernova data using information field theory. \texttt{charm2} improves over \texttt{charm1} \citep{Porqueres_2017} by providing more accurate uncertainties and a better treatment of the nonlinearities in the data, both of which are achieved by using the geometric variational inference algorithm \citep{Frank_2021}. \texttt{charm2} also uses a flexible Bayesian hierarchical model that permits inference of central quantities, such as the mean background cosmology and the statistical power spectrum, jointly with the quantities of interest. This ensures that \texttt{charm2} is robust against biases arising from assuming a cosmological model. 

We tested the effect of different initializations of the fluctuation parameter $b_0$, which regulate the expected deviations from the mean background cosmology. We demonstrated that the analyses of Union2.1 and Pantheon+ data were robust against the initial choice of $b_0$ and the inferred reconstructions are compatible with flat $\Lambda$CDM. 

The Union2.1 catalog, which assumes $H_0 = 70\:\mathrm{km/s/Mpc}$ for data calibration, results in a reconstruction mostly compatible with Planck2018 \citep{Planck_2018}, whereas the Pantheon+ reconstruction deviates significantly from Planck2018 at lower redshifts, a manifestation of the Hubble tension as discussed in \cite{Riess_2022}.

The reconstruction from DESY5 data showed a significant deviation from flat $\Lambda$CDM benchmark fields at  $x\gtrsim 0.45$ ($z\gtrsim 0.57$) for initial fluctuation values of $b_0 \gtrsim 0.1$. Through a numerical experiment using synthetic reconstructions, we ruled out the possibility that this deviation could be explained by a systematic offset in the distance moduli in the redshift range suggested by \cite{Efstathiou2024}. We showed that our DESY5 reconstruction is compatible with an evolving dark energy signal. Applying $\texttt{charm2}$ with $b_0=0.6$, which allows for larger deviations from flat $\Lambda\mathrm{CDM}$, to the updated DESY5-Dovekie sample, resulted in a shift of the posterior mean field towards lower energy density at high redshift, effectively relaxing the field compared to the original DESY5 reconstruction. This suggests that the nonparametric model was affected by systematics present in DESY5 that are mitigated in DESY5-Dovekie.

Using the evidence lower bound as a proxy for the evidence, we find that flat $\Lambda\mathrm{CDM}$ is consistently preferred over our nonparametric reconstructions. Accordingly, the supernova samples analyzed here do not provide evidence for non-flat-$\Lambda\mathrm{CDM}$ features that would be captured by a nonparametric model. However, we demonstrated through a synthetic data example that at current SNIa noise levels, the evidence lower bound and, by extension, the evidence might point to flat $\Lambda\mathrm{CDM}$ although the ground truth is evolving dark energy. In this example, the evidence lower bound correctly preferred the nonparametric model only when the noise level was reduced by a factor of $\gtrsim 7$.

In summary, we have developed a nonparametric, model-agnostic method to infer the cosmic history from SNIa data. This method improves the set of tools we can use to rationally question and investigate the impact of assumptions on data analysis. 

Possible extensions of the \texttt{charm2} method are the inclusion of other data sets, such as Cepheids, and the use of the SNe spatial distribution to constrain anisotropies of the expansion rate.

\begin{acknowledgements}
    The authors thank Hanieh Zandinejad for helpful discussions and Lena Tauber for support in preparing the manuscript. M.G. acknowledges support from the European Union-funded project $\texttt{mw-atlas}$ under grant agreement No. 101166905.
\end{acknowledgements}


\bibliographystyle{aa} 
\bibliography{references} 

\appendix 

\section{\texttt{geoVI} transform and standardization}\label{app: Working principle of geoVI}

Any variable $s\hookleftarrow \mathcal{G}(s,S)$ can be reparameterized in terms of a standard normal "latent variable" $\xi_s \hookleftarrow \mathcal{G}(\xi_s, \mathbf{1})$. This shifts the complexity of the correlation structure within the probability distribution to the mapping between $s$ and $\xi_s$, allowing to construct hierarchical prior models (see Fig. \ref{fig: Computational graph of Correlated Field Model}) more easily.  For a Gaussian field $s$, the mapping is achieved by considering an amplitude operator $A = F^{\dagger}\sqrt{\widehat{p}_s}F$, where $F$ is the Fourier transform, and $\widehat{p}_s$ is an operator with the signal power spectrum $p_s$ as its diagonal: the field $s$ can then be written as 
\begin{equation}\label{eq: app: signal from amplitude operator}
    s=A\xi_s,
\end{equation}
which is a zero-mean, Gaussian variable with the covariance $\langle ss^{\dagger}\rangle_{(s)}=A\langle \xi_s \xi_s^{\dagger}\rangle_{(\xi_s)}A^{\dagger}=AA^{\dagger}=F^{\dagger} (\sqrt{\widehat{p}_s})^2F = S$. Note that the power spectrum $p_s$ and by extension the amplitude operator $A$, are themselves built from standard normal variables (cf. Fig. \ref{fig: Computational graph of Correlated Field Model}). All standard normal variables involved in generating the model are collected into a single latent parameter vector $\xi$. To find an approximation to the posterior $p(\xi):=\mathcal{P}(\xi\vert d)$, the KL-divergence between the posterior and a parameterized trial distribution $\mathcal{Q}_\theta (\xi)$ is minimized. The KL-divergence reads
\begin{equation}\label{eq: KL divergence}
    \mathcal{D}_\mathrm{KL}(\mathcal{Q}_\theta (\xi) \:\Vert\: p(\xi)) = \int \mathrm{D}_\xi\: \mathcal{Q}_\theta (\xi) \mathrm{ln}\bigg( \frac{\mathcal{Q}_\theta (\xi)}{p(\xi)}\bigg),
\end{equation}
where $\mathrm{D}_\xi$ denotes an integration over all possible field configurations of $\xi$. In order to ensure that a Gaussian distribution, completely parameterized through a variable $\mathcal{\theta}$,
\begin{equation*}
    \mathcal{Q}_{\theta}(\xi) = \mathcal{G}(\theta,\Xi(\theta)),
\end{equation*}
is a good approximating distribution, \texttt{geoVI} performs a coordinate transformation that aims to flatten out potential curvature in the posterior distribution. To extract the necessary geometrical information from the posterior, a metric $\mathcal{M}$ is introduced:  
\begin{equation}\label{eq: app: Metric in geoVI}
    \mathcal{M} =\mathrm{F}(\theta) + \mathbf{1},
\end{equation}
with the Fisher Information Metric 
\begin{equation}
    \mathrm{F}(\theta) = \bigg\langle \frac{\partial \mathcal{H}(d\mid \xi)}{\partial \xi} \frac{\partial \mathcal{H}(d\mid \xi)}{\partial \xi^{\dagger}} \bigg\rangle_{(d\mid \xi=\theta)}.
\end{equation}
A coordinate transformation $g(\xi)$ is desired through which the metric, Eq. \eqref{eq: app: Metric in geoVI}, becomes "diagonal" in the following sense:
\begin{equation}
    \mathcal{M}\stackrel{!}{=} \bigg( \frac{\partial g}{\partial \xi}\bigg)^{\dagger} \mathbf{1} \bigg( \frac{\partial g}{\partial \xi}\bigg).
\end{equation}
\cite{Frank_2021} show that this transformation does not exist globally, but an expression that approximates the metric around an expansion $\bar{\xi}$ does:
\begin{equation}\label{eq: }
    g(\xi, \bar{\xi}) = \sqrt{\bar{\mathcal{M}}}^{-1}\bigg ( \xi - \bar{\xi} + \bigg(\frac{\partial y}{\partial \xi} \bigg|_{\bar{\xi}} \bigg)^{\mathrm{T}} (y(\xi)-y(\bar{\xi}))\bigg)
\end{equation}
where $\bar{\mathcal{M}}$ is the local metric at the expansion point and $y(\xi)$ is a function through which the Fisher Information Metric also becomes diagonal: 
\begin{equation}\label{eq: app: FIM needs to be diagonal}
   \mathrm{F}(\theta) \stackrel{!}{=} \bigg(\frac{\partial y}{\partial \xi}\bigg)^{\dagger} \mathbf{1}\bigg(\frac{\partial y}{\partial \xi}\bigg) \bigg|_{\xi=\theta}.
\end{equation}
For a Gaussian likelihood, the Fisher Information Metric can be calculated as 
\begin{equation}\label{eq: app: FIM for Gaussian likelihood}
    \mathrm{F}(\theta) = \bigg(\frac{\partial s^{\prime}}{\partial \xi}\bigg)^{\dagger} N^{-1}\bigg(\frac{\partial s^{\prime}}{\partial \xi}\bigg) \bigg|_{\xi=\theta},
\end{equation}
where $s^{\prime}$ is the noiseless data, $s^{\prime}:=R(s(\xi))$. Finally, a comparison between Eq. \eqref{eq: app: FIM needs to be diagonal} and Eq. \eqref{eq: app: FIM for Gaussian likelihood} suggests 
\begin{equation*}
    y = \sqrt{N^{-1}}s^{\prime},
\end{equation*}
where the noise covariance $N$ is provided together with the distance moduli analyzed in this work. To take the matrix square root, we diagonalize the inverse noise covariance matrix and take the square root of the diagonal elements: 
\begin{equation}
    \sqrt{N^{-1}} = O \sqrt{\Lambda} O^{\dagger},
\end{equation}
where $\sqrt{\Lambda}$ is a diagonal matrix containing the square root of the eigenvalues of $N^{-1}$ and $O$ is the matrix consisting out of the unit-length eigenvectors of $N^{-1}$. 
 
\section{Construction of reference fields}\label{sec: app: Construction of reference fields}

Fig. \ref{fig: Comparison Fields for flat LCDM} shows the signal $s(x)$ for the cosmological parameters (cf. Table \ref{table: Adopted values of H0 and Omega_m for comparison of reconstructions to flat LCDM model}) found by the  \cite{Planck_2018} ($s_{\text{CMB}}$) and \cite{Pantheon_2022} ($s_{\text{SN}}$). These fields are parametric flat $\Lambda$CDM models, i.e.
\begin{equation}\label{eq: Flat LCDM Friedmann Equation}
    \frac{H(a)^2}{H_0^2} = \Omega_{\Lambda}+\Omega_ma^{-3}.
\end{equation}
This corresponds to the signal (see Eq. \ref{eq: First Friedmann Equation}) of
\begin{equation}\label{eq: Translation of flat lcdm into signal field}
    s(x) = \mathrm{ln}\bigg(\frac{3}{8\pi G}H_0^2 (1+\Omega_m(e^{3x}-1))\bigg).
\end{equation}
 The comparison of $s_{\text{CMB}}$ and $s_{\text{SN}}$ illustrates the impact of different $H_0$ values on the signal's shape and amplitude. The $1\sigma$ regions in Fig. \ref{fig: Comparison Fields for flat LCDM} are calculated through Gaussian propagation of the errors in Table \ref{table: Adopted values of H0 and Omega_m for comparison of reconstructions to flat LCDM model} following to Eq. \eqref{eq: Translation of flat lcdm into signal field}. 

\begin{table}[h]
    \centering
    \setlength{\tabcolsep}{12pt} 
    \caption{Adopted values of $H_0$ and $\Omega_m$ from Pantheon+SH0ES and Planck2018 cosmological results, used for constructing flat $\Lambda$CDM reference fields to compare with \texttt{charm1} and \texttt{charm2} reconstructions.}
    \label{table: Adopted values of H0 and Omega_m for comparison of reconstructions to flat LCDM model}
    \begin{tabularx}{0.5\textwidth}{l c c}  
        \toprule
        & \makecell{\textbf{Pantheon+SH0ES} \\ \cite{Pantheon_2022}} 
        & \makecell{\textbf{Planck2018} \\ \cite{Planck_2018}} \\
        \midrule
        $\hat{H}_0$ & $73.60 \pm 1.1$ & $67.36 \pm 0.54$ \\
        $\Omega_m$ & $0.334 \pm 0.018$ & $0.3153 \pm 0.0073$ \\
        \bottomrule
    \end{tabularx}
\end{table}

\section{Converting \texttt{charm1} signal fields}\label{app: Note on converting charm1 fields}

For a meaningful comparison of the reconstructions between \texttt{charm1} and \texttt{charm2}, we need to make adjustments due to slightly different signal definitions.  

We translate the \texttt{charm1} signal field, denoted by $s_{\text{\texttt{charm1}, before}}:=\mathrm{ln}(\rho /\rho_{c,0})$, to the corresponding \texttt{charm2} signal field satisfying the new definitions Eq. \eqref{eq: Signal definition v2} and Eq. \eqref{eq: energy density normalization} through
\begin{equation}\label{eq: Translation from charm1 to new definition}
    s_{\text{\texttt{charm1}, after}} = \mathrm{ln}((\rho_{c,0}/\rho_0) e^{s_{\text{\texttt{charm1}, before}}}),
\end{equation}
where $\rho_0$ is the reference energy density introduced in this work, Eq. \eqref{eq: energy density normalization}.
The original \texttt{charm1} reconstruction used only the \citetalias{Suzuki_2012} dataset, which, at that time, was the latest available. 

\section{Details on the evidence lower bound}\label{app: Details on the evidence lower bound}

In an information field theory context, the ELBO has been previously successfully applied to the problem of inferring causal directions between random variables \citep{GuardianiCovid}. It is illustrative to explicitly present the ELBO terms underlying that work by writing $q(\xi)$ for the approximating distribution, $p(\xi)$ for the posterior and $\mathcal{H}_\mathcal{P}$ for the negative natural logarithm of probability $\mathcal{P}$: then, Eq. \eqref{eq: KL divergence} can be rewritten as 
\begin{equation}\label{eq: Rewriting KL to show log evidence and ELBO}
    \mathrm{ln}\:\mathcal{P}(d) = \mathcal{D}_{\mathrm{KL}}(q(\xi)\:\Vert\:p(\xi)) + \langle \mathcal{H}_q(\xi) \rangle_q  - \langle  \mathcal{H}_p(d,\xi) \rangle_q.
\end{equation}
The $\mathrm{ELBO}$ is defined as 
\begin{equation}\label{eq: ELBO intermediate step}
    \mathrm{ELBO} := \langle \mathcal{H}_q(\xi) \rangle_q  - \langle  \mathcal{H}_p(d,\xi) \rangle_q.
\end{equation}
The first term, the average of the approximating distribution's negative logarithm, can be calculated to be 
\begin{equation}
    \langle \mathcal{H}_q(\xi) \rangle_q  = \frac{1}{2}\mathrm{ln}\vert 2\pi e \Xi \vert , 
\end{equation}
where $e$ is Euler's number and $\Xi$ is the posterior covariance of $q$. The second term can be split into a likelihood and prior term: 
\begin{equation}
    \langle  \mathcal{H}_p(d,\xi) \rangle_q = \langle  \mathcal{H}_p(d\vert \xi) \rangle_q+\langle  \mathcal{H}(\xi) \rangle_q.
\end{equation}
Thanks to the standardization procedure, $\mathcal{P}(\xi)$ is by construction standard normal and the second term can be calculated as 
\begin{equation}
    \langle  \mathcal{H}(\xi) \rangle_q = \frac{1}{2}\bigg(N\mathrm{ln}(2\pi) + \mathrm{Tr}(\Xi) + \bar{\xi}^\dagger \bar{\xi}\bigg),
\end{equation}
where $\bar{\xi}$ denotes the posterior mean of $q$. Collecting all the terms and using the identity $\mathrm{ln} \:\mathrm{det}( \Xi) = \mathrm{Tr}\:\mathrm{ln}(\Xi) $, one finds
\begin{equation}
    \mathrm{ELBO} =  -\langle \mathcal{H}(d\vert \xi)\rangle_q + \frac{1}{2}\mathrm{Tr}\:\mathrm{ln}(\Lambda_\Xi) - \frac{1}{2}\mathrm{Tr}(\Lambda_\Xi-\mathbf{1}) - \frac{1}{2}\bar{\xi}^\dagger \bar{\xi} \label{eq: terms of ELBO} 
\end{equation}
Here, $\Lambda_\Xi$ denotes the diagonalized matrix of the posterior covariance $\Xi$, needed for the calculation of the log determinant. Note that the ELBO decreases with an increase in the Occam's Razor term, $\frac{1}{2}\bar{\xi}^\dagger \bar{\xi}$: the more degrees of freedom need to deviate from their prior mean of $0$ in latent space to explain the data, the less favored the model is.

Comparing Eq. \eqref{eq: Rewriting KL to show log evidence and ELBO} and Eq. \eqref{eq: ELBO intermediate step}, it is clear that the $\mathrm{ELBO}$ calculated this way provides a lower bound on the log-evidence: 
\begin{equation}
    \mathrm{ln}\: \mathcal{P}(d) \geq \mathrm{ELBO}.
\end{equation}

\end{document}